\documentclass{article}
\usepackage[utf8x]{inputenc}
\usepackage[a4paper]{meta-donnees}		
\usepackage{graphicx}	
\usepackage{hyperref}							
\usepackage[final]{pdfpages}
\usepackage{natbib} 
\usepackage{algorithm}
\usepackage{algorithmic}

\usepackage{booktabs}% for tabulars

\usepackage{amsmath, amsthm, amssymb}
\usepackage{color}
\usepackage{epsfig}
\usepackage{verbatim} % for commenting sections
\usepackage{caption}
\usepackage{subcaption}
\usepackage{enumerate}
\usepackage[toc,page]{appendix} 

\title{A Model for Scaling in Firms' Size and Growth Rate Distribution}
\author{Cornelia Metzig$^\dagger$ and Mirta B. Gordon$^\dagger$,$^*$ \\
$^\dagger$Universit\'e Joseph Fourier Grenoble 1 / $^*$CNRS\\ Laboratoire LIG \\
\texttt{cornelia.metzig@imag.fr}, \texttt{mirta.gordon@imag.fr}
}
\date{}

\begin{document}
\maketitle
%\tableofcontents
\begin{abstract}
 We introduce a simple agent-based model which allows us to analyze three stylized facts: a fat-tailed size distribution of companies, a `tent-shaped' growth rate distribution, the scaling relation of the growth rate variance with firm size, and the causality between them. This is achieved under the simple hypothesis that firms compete for a scarce quantity (either aggregate demand or workforce) which is allocated probabilistically. The model allows us to relate size and growth rate distributions. We compare the results of our model to simulations with other scaling relationships, and to similar models and relate it to existing theory. Effects arising from binning data are discussed.
\end{abstract}

\section{Introduction}\label{sec:related}
Economic growth processes have been the object of active research since the ground-laying work of R. Gibrat \cite{gibrat1931inegalites} who described growth as a multiplicative stochastic process. By assuming that growth rates are independent and identically distributed random variables, and by studying the time evolution of the system, he obtained a lognormal distribution of company sizes, i.e. a heavy-tailed distribution. The topic has since then attracted much interest \cite{sutton1997gibrat}. Recent empirical studies suggest that the firm size distribution follows a Zipf (power law) distribution \cite{bottazzi2013zipf, axtell2001zipf,okuyama1999zipf}. Various different models exist for the formation of such distributions with power-law tails. In the context of firm growth, the well-known model by H. Simon \cite{simon1955class} explains a power law for firm size distribution based on a process introduced by Yule \cite{newman2005power}. In his model, the number of firms is constantly growing in the system, and indvidual firms cannot shrink, i.e. they grow continuously. The exponent of the power law depends on the frequency of new firms, and approaches $1$ if this frequency is low. However, the requirement that the system has to grow continuously for the formation of a power law limits its applicability \cite{newman2005power, gabaix2008power}. For systems of constant global size, models exist that explain the formation of fat tails by multiplicative stochastic processes. Although for such systems with multiplicative noise, no result of the generality of the central limit theorem exists \cite{mondani2012statistical}, it has been shown that the stationary distribution has a power law tail in the lowest order approximation, if in addition additive noise is present\cite{takayasu1997stable,biro2005power,sornette1998multiplicative,levy1996power}. 
 
More recently, Stanley et al. \cite{stanley1996scaling} uncovered two empirical features that an accurate theory on firm growth should explain. The first is that the growth rate frequencies exhibit exponential (Laplacian) decay, i.e. it gives rise to a tent shape in logarithmic scale (see also \cite{bottazzi2006explaining, alfarano2012statistical, erlingsson2013distributional}). The second is that the variance of the growth rate scales with company size $n$ as $\sigma(n)\propto n^{-\beta}$. This means that the simple assumptions by Gibrat and others, who assume multiplicative noise to be independent of firm size, are at odds with the data. The empirically determined values of $\beta $ (typically $\approx$ $0.2$) depend on the studied system. A number of papers have provided further evidence of these two findings in various growth processes: firm growth \cite{stanley1996scaling} ($\beta =0.15$), \cite{nunes1997scaling} ($\beta$ = $0.18$), \cite{fu2005growth} ($\beta$ = $0.28$), \cite{schwarzkopf2010explanation} ($\beta = 0.3$) (the latter authors also consider bird populations and mutual funds), a country's GDP growth \cite{lee1998universal}  ($\beta$ = $0.15$), citations in scientific journals  \cite{picoli2006scaling}  ($\beta$ = $0.22$) and the growth rate of crime \cite{alves2013scaling}  ($\beta$ = $0.36$). The multitude of examples suggest that the process generating a tent-shaped growth rate distribution, a scaling exponent $\beta$ $\neq$ $0$ for the growth rate standard deviation, and a fat tailed size distribution is simple and universal.

A number of sophisticated models giving rise to a tent shaped growth rate distribution have been proposed. They follow different approaches. Bottazzi and Secchi \cite{bottazzi2006explaining} predict a tent-shaped (Laplacian) growth rate distribution as being the result of a number of abstract shocks drawn from a Polya urn, without addressing the question of the standard deviation's scaling exponent. In order to obtain the scaling of the standard deviation, many models assume that firms have an internal structure, i.e. they are composed of subunits \cite{wyart2003statistical, fu2005growth, takayasu2013generalized}.
In Schwarzkopf et al.\cite{schwarzkopf2010explanation}, the  probability that the firms' subunits reproduce themselves follows a power law. As a result, the aggregated growth rate distribution is tent-shaped with a power law decay: it is not a collective phenomenon but holds at the individual level. Another interesting model \cite{picoli2006scaling} assumes that the growth rate variance depends on the size of the elements (which are citations in their case), and numerically obtains a fat-tailed size distribution. The tent-shaped growth rate distribution is however not a result of the model but instead a hypothesis at the individual level.

Most models explaining the tent-shaped growth rate distribution and the variance scaling relation do not attempt to simultaneously explain the formation of the fat-tailed firm size distribution. Rather, existing models for power law tails via multiplicative noise assume the growth rate to be independent of the firms' size, which seems in conflict with a scaling exponent $\beta$ $\neq$ $0$. However, there is empirical evidence for both a power law firm size distribution and a scaling exponent $\beta > 0$ for the standard deviation.

%We address this issue in this paper,
This issue is addressed in this work, both theoretically and numerically with a simple agent-based model comprising firms and employees which conserves the flow of funds. A distinction is made between collective phenomena at the firm level and at the level of the macroeconomy.  In contrast to the models cited above, we investigate the hypothesis that the same process accounts for the tent-shaped growth rate distribution, its standard deviation's scaling exponent $\beta$ $>$ $0$ and the fat-tailed size distribution of firms.

This paper is organized as follows. In section \ref{sec:model}, the model is introcuced. In section \ref{sec:power_langevin}, existing theory is recalled. In section \ref{sec:theo_analysis_model}, the model and its growth rate distribution are described theoretically.  In section \ref{sec:num_results_all} we provide numerical results of a heavy tailed size distribution, which is our first main result. Then we present the growth rate distribution of the model, which is our second main result, and discuss some consequences of binning the data. In section \ref{sec:scaling_relation} we compare empirically found scaling exponents to our model and point out possible extensions of it. In section \ref{sec:discussion}, we discuss a theoretical aspect of the model and compare it to further literature, then we conclude and point out extensions. 

\section{The model}\label{sec:model}
In this section, a simple firm growth model is introduced, and subsequently analyzed in the following sections. The model is composed of firms and workers, $N_f$ and $N_w$, which are both constant numbers, and models a simplified production cycle. Depending on the considered setting, firms compete either for purchasing power or for workforce. This introduces a stochastic element and accounts for the growth dynamics. Firms are characterized by an expected profit margin $\mu=$  (sales-expenses)/expenses, which allows them to determine the necessary amount of workers $n$ in order to produce a given quantity $q$ of commodity goods. Goods are sold at a fixed price $p$, workers earn a wage $w$. The expected margin of a firm $i$ is defined as 
\begin{equation}
 \mu_i=\frac{q_i\,p-n_i\,w}{n_i\,w}\,.
\label{eq:mu_expected}
\end{equation}
\begin{comment}
 This allows firms to calculate the number of workers $\hat n_i$
\begin{equation}
 \hat n_i=\hat q_i \frac{p}{w}\left(\frac{1}{1+\mu_i}\right)\,.
\label{eq:n_hat}
\end{equation}
Inversely, the produced quantity of a firm with $n_i$ workers is
\end{comment}
This relation allows a firm to calculate the neccessary number of workers in order to produce a certain amount of goods,
\begin{equation}
 \hat n_i=\hat q_i \frac{p}{w}\left(\frac{1}{1+\mu_i}\right)\,,
\label{eq:n_hat}
\end{equation}
 and inversely, to calculate how many goods it can produce with a given number of workers: 
\begin{equation}
 q_i=n_i\frac{w}{p}(1+\mu_i)
\label{eq:q_i}
\end{equation}

The size of a firm $i$ is defined as the number of its workers. In the following, $\hat q_{i,t}$ is the intended production quantity of firms, $q_{i,t}$ the actual produced quantity, and $q^s_{i,t}$ the sold quantity at iteration $t$. At each iteration, firms $i$ hire $n_i$ workers in a job market in order to produce a quantity of $q_i$ goods, which are then sold in a goods market. Firms adjust the quantity $\hat q_i$ they indend to produce at time $t$ based on the previously produced quantity $q_{i,t-1}$, and on their last realized profits:
\begin{equation}
 \hat q_{i,t} =q_{i,t-1}(1+\mu^n_{i,t-1})\,,
\label{eq:q_hat}
\end{equation}
where $\mu^n_{i,t-1}$ is the net realized profit margin 
\begin{equation}
\mu^n_{i,t-1}=\frac{q_i^s\,p-n_i\,w}{n_i\,w}\,.
\end{equation}
It may be smaller than the expected margin $\mu_i$. In order to be able to produce this quantity, firms offer $\hat n_i$ jobs (\ref{eq:n_hat}). The number of job offers $\hat n_i$ which each firm posts in the job market is
\begin{equation}
 \hat n_i=\hat q_i\,(1+\mu_{i})\frac{p}{w}\,.
\label{eq:n_hat_inserted}
\end{equation}
The job market then opens, where open positions and available workers are matched at random. Every open position has the same probability of being filled by a worker, or in the case of too few jobs, every worker has the same chance of being hired. After the job market, production takes place, where firms produce $q_i$ goods according to equation (\ref{eq:q_i}). These goods are put in a goods market, which follows the same algorithm as the job market: every produced good has the same chance of being sold, or in the case of shortage of goods, every demand has the same chance of being satisfied.
This hypothesis, which is equivalent to the microcanonical ensemble in statistical physics, is also used elsewhere in the context of growth processes, albeit for internal firm structure \cite{sutton2002variance, wyart2003statistical}\footnote{The concept of a statistical equilibrium in a market has also been applied to markets with more general hypotheses \cite{foley1994statistical}, which applies to the markets in this model as a special case.}. 

Throughout this paper, the expected profits are assumed to be the same for all firms \footnote{A scenario with heterogeneous $\mu_i$ of firms is discussed in \cite{metzig2013heterogeneous}.}. The following two cases are discussed separately: 
\begin{enumerate}[(i)]
\item both workers and firms are consumers, who spend their salaries and their profits in the goods market
\item only workers spend their salaries. 
\end{enumerate}
Firms compete for two limited resources, workforce and purchasing power of customers. In case (i), purchasing power is sufficient and the limited resource is workforce. In case (ii) only workers are consumers, and firms compete for limited purchasing power. Since the job market and the goods market are based on the same algorithm, these two settings lead to the same evolution of firm size distribution and growth rate distribution, but for clarity they are presented separately in subsections \ref{sec:job_market} and \ref{sec:goods_market}.

In case (i), aggregate demand is sufficient, firms sell their entire production, i.e. $q^s_{i,t}=q_{i,t}$ and $\mu^n_{i,t}=\mu_i$, and demand a discrete quantity of $\hat n_i$ workers, proportional to their size
\begin{equation}
 \hat n_{i,t}= n_{i,t-1}(1+\mu)\,.
\label{eq:n_hat_case1}
\end{equation}
 Since $\mu>0$, firms attempt to increase their size, and the stationary state of the system corresponds to full employment. In case (ii), not all goods $q_{i,t}$ are sold, i.e. $q^s_{i,t}<q_{i,t}$ and $\mu^n_{i,t}<\mu_i$. As a consequence, the same number of workers will be hired as in the previous interation, and the same total number $Q_t$ of goods will be produced. In that case, firms demand less workers than available, and do not compete for workforce.

\subsection{The job market}\label{sec:job_market}
In the job market, workers do not stay at their firm but are newly placed at every iteration\footnote{Similar results would hold if only a (constant) fraction of workers of each company were newly placed}.
In case (i), there is a workforce shortage, and the allocation of workers gives rise to interesting dynamics: since the $N_w\,(1+\mu)$ open positions are covered at random with $N_w$ workers, the actual number $n_{i,t+1}$ of employees hired by a firm $i$ at time $t+1$, is likely to be smaller than their job offer $\hat n_{i,t}=n_{i,t}(1+\mu)$. 

It can even be smaller than the number of employees $n_{i,t}$ in the preceding period, which includes a situation where firms may receive no worker at all and vanish. The number of active firms would decrease continuously, and workers would eventually accumulate in a monopoly, which is avoided by the introduction of new firms. To maintain $N_f$ constant, extinct firms are replaced by new ones\footnote{It is also possible to re-insert firms at a constant rate, in which case the level of active firms will become stationary after some time. In this chapter, a strictly constant number of active firms is used, since this guarantees that all presented systems will be the same size.}, initialized with a number of workers $n^{new}_{i,t}$ drawn from a distribution $\mathcal{F}(n^{new})$. New firms contribute to the total demand for workforce in the next period with the quantity $\hat n^{new}_t=(1+\mu) n^{new}_{i,t}$. 
 Analytically, the matching process in a market is described by a multinomial probability distribution. A simpler description of the evolution of the system is obtained if the number of workers is assumed to be conserved only on average: $\langle \sum_i n_{i,t} \rangle=N_w$. 
The probability for an open position to be filled then becomes
\begin{equation}
p = \frac{N_w}{\sum_i \hat{n}_i} \,,
\label{eq:p}
\end{equation}
where $\sum_i \hat{n}_i= (1+\mu)\,N_w + \sum_i {n}^{new}_i$. Since $\sum_i {n}^{new}_i \ll N_w$, the probability of a position to be filled is approximately $p\approx (1+\mu)^{-1}$.
%Under such conditions 
The probability for a firm of size $n_i$ to receive $k_i$ workers in the next period is given by the binomial distribution
\begin{equation}
 P(k_i | n_i)=\binom{\hat n_i}{k_i}p^{k_i}(1-p)^{\hat n_i-k_i} \, ,
\label{eq:binomial_dist}
\end{equation}
with mean $\left\langle n_{i,t}\right\rangle =\hat{n}_{i,t}\, p = n_{i,t-1}$, which is precisely the number of workers of firm $i$ at the previous time step. The variance is $\hat {n}_i \,p\,(1-p)=n_i \mu / (1+\mu)^2$.
With probability $P(0 | n_i)=(1-p)^{\hat n_i}$ a firm does not receive any workers and disappears.

For large $n_i$, the binomial distributions may be approximated by Gaussian distributions, whose variance exhibits the same $n$-dependence. Alternatively, another implementation uses a different rounding method by which firms determine their job offer, such that the Gaussian distribution is a good approximation even for small firms. It can be seen as growth of entirely independent subunits, where subunits are jobs (see appendix \ref{sec:rounding_joboffer}). For both the alternative method and for equation (\ref{eq:binomial_dist}), the Gaussian approximation of the probability for a firm of size $n_{i, t}$ to reach size $n_{i,t+1}$ is written as
\begin{equation}
 \mathcal{P}(n_{i,t+1}=k_{i}|n_{i,t})=\frac{1}{\sigma_{i,t}\sqrt{2\pi}} e^{-\frac{1}{2}\left(\frac {k_{i}-n_{i,t}}{\sigma_{i,t}}\right)^{2}}\,,
\label{eq:gaussian_initial}
\end{equation}
where the mean has been replaced by its value $n_{i,t}$, and $\sigma_{i,t}^2=n_{i,t}\frac{\mu}{(1+\mu)^2}=c\,n_{i,t}$. If the growth rate of a firm is defined as
\begin{equation}
g_{i,t}= \frac{n_{i,t+1}}{n_{i,t}}\,,
\label{eq:growth_rate}
\end{equation}
equation (\ref{eq:gaussian_initial}) yields for the growth rate probability density $\cal{G}$ (dropping the index $t$):
\begin{equation}
 \mathcal{G}(g_{i}|n_{i})=\sqrt{\frac{n_{i}}{2\pi\,c}} e^{-\frac{1}{2} \frac{n_i}{c}\, (g_{i}-1)^{2}} \,,
\label{eq:gaussian_growthrate_independent}
\end{equation}
where $c=\mu/(1+\mu)^2$ for the binomial approximation\footnote{For the rounding method detailed in appendix \ref{sec:rounding_joboffer}, this constant is $\frac{2\mu}{(1+\mu)^2}$}.
Thus, in the present model, the scaling exponent $\beta $ of the growth rate's standard deviation is defined through
\begin{equation}\sigma(n)\propto n^{-\beta}\,,
\label{eq:scaling_relation}
\end{equation}
and has the value $\beta$ = $0.5$. This value for $\beta$ is a general feature of models that explain firm growth as being the sum of the growth of independent subunits. In other published models, subunits often represent the sectors in which the firm is active
\cite{nunes1997scaling,sutton2002variance,wyart2003statistical}; in this model these are jobs. Other values for $\beta $ and the corresponding empirical evidence are addressed in section \ref{sec:scaling_relation}. 
\subsection{The goods market}\label{sec:goods_market}
In scenario (ii), firms do not spend their profits in the goods market. In that case, the aggregate demand $D=\sum_j d_j$ consists only of the wages which are paid to the $j$ employees, and is a scarce quantity, whereas in scenario (i) aggregate demand matched aggregate offer and the goods market had no importance for the dynamics. In scenario (ii), a quantity $\sum_i q_{i,t}\,p$ is produced, but the demand, i.e. the overall wages paid to workers are
\begin{equation}
D_t=\sum_i n_{i,t}\,w=\sum_i q_{i,t}\,p\,\frac{1}{1+\mu}\,.
\end{equation} 
It is clear that the aggregate sales $\sum_i q^s_{i,t}\,p$ is smaller than the production $Q_t=\sum_i q_{i,t}\,p$, since in this scenario workers are the only consumers. Then, the probability for a produced good to be sold becomes
\begin{equation}
v =\frac{D_t}{Q_t}= \frac{\sum_i q_{i,t}\,p\,\frac{1}{1+\mu}}{\sum_i q_{i,t}\,p} \,=\frac{1}{1+\mu}\,,
\label{eq:s}
\end{equation}
which is the analogon to equation (\ref{eq:p}). Unsold goods are lost, they cannot be stored and put in the market in following iterations. Since every good has the same chance of being sold, the allocation of demand also follows a binomial distribution
\begin{equation}
 V(q^s_i | q_i)=\binom{ q_i}{q^s_i}v^{q^s_i}(1-v)^{ q_i- q_i^s} \,.
\label{eq:binomial_dist_goods}
\end{equation}
Again, this binomial distribution can be approximated with a Gaussian for large $q$. It becomes
\begin{equation}
 \mathcal{V}(q^s_{i,t+1}=k_{i}| q_{i,t})=\frac{1}{\sigma_i\sqrt{2\pi}} e^{-\frac{1}{2}\left(\frac { k_{i}-\hat q_{i,t}}{\sigma_{i}}\right)^{2}}\,.
\label{eq:gaussian_initial_goods}
\end{equation}
On average, each firm sells a quantity $ \left\langle q^s_{i,t} \right\rangle = q_{i}\, v = \frac{w}{p}n_{i,t}$. Therefore, their average realized profit is $\left\langle \mu_{i,t}^{net}\right\rangle =\frac{\left\langle q^s_{i,t}p\right\rangle-n_{i,t}w}{n_{i,t}w}=0$. In the next iteration, firms will demand $\left\langle\hat n_{i,t+1}\right\rangle=n_{i,t}$, and since there is no competition of workforce, $\left\langle n_{i,t+1}\right\rangle=n_{i,t}$. The average quantity of sold goods can be expressed in terms of the previously sold quantity: $ \left\langle q^s_{i,t} \right\rangle = q_{i}\, v = \frac{w}{p}n_{i,t}= \frac{w}{p}\left\langle n_{i,t-1} (1+\mu_{i,t-1}^{net})\right\rangle=\frac{w}{p}\left\langle n_{i,t-1}(\frac{q^s_{i,t-1}}{n_i,t-1})\right\rangle=\left\langle q^s_{i,t-1}\right\rangle$. Thus, on average, firms stay constant in size, both measured in terms of employees, and in terms of sales. The growth rate can be written as 
\begin{equation}
g_{i,t}= \frac{n_{i,t+1}}{n_{i,t}}
\label{eq:growth_rate}
\end{equation}
or equivalently as
\begin{equation}
g^{sales}_{i,t-1}= \frac{q^s_{i,t}}{q^s_{i,t-1}}\,.
\label{eq:growth_rate}
\end{equation}
%\begin{equation}
%g_{i,t}= \frac{q^s_{i,t+1}}{q^s_{i,t}}
%\label{eq:growth_rate}
%\end{equation}
This yields the growth rate probability density
\begin{equation}
 \mathcal{G}(g_{i}|q^s_{i})=\sqrt{\frac{q^s_{i}}{2\pi\,c}} e^{-\frac{1}{2} \frac{q^s_i}{c}\, (g_{i}-1)^{2}} \,,
\label{eq:gaussian_growthrate_independent_goods}
\end{equation}
%\begin{equation}
% \mathcal{G}(g_{i}|q^s_{i})=\sqrt{\frac{q^s_{i}}{2\pi\,c}} e^{-\frac{1}{2} \frac{q^s_i}{c}\, (g_{i}-1)^{2}} \,,
%\label{eq:gaussian_growthrate_independent_goods}
%\end{equation}
with $c=\frac{\mu}{(1+\mu)^2}$, where the standard deviation of the growth rate scales as
\begin{equation}\sigma(q)\propto q^{-\beta}\,,
\label{eq:scaling_relation_goods}
\end{equation}
with $\beta =0.5$, as in the job market competition scenario.

\section{Preliminaries for analyzing the dynamics}\label{sec:power_langevin}
In the following, the dynamics of the model is analyzed. Iteration of scenarios (i) or (ii) each result in the same fat-tailed size distribution that can be fitted by a power law
\begin{equation}
 P(x\geq x')=x^{-\alpha}
\end{equation}
 which is however not a Zipf law with exponent $\alpha=1$. 
In order to be able to describe it, some existing theory on the formation of power laws is recalled.   

\paragraph{The Langevin equation.}
The Langevin equation was introduced by P. Langevin in 1908 in order to describe Brownian motion of particles in a fluid \cite{langevin1908theory}. 
\begin{equation}
 n_{t+1}=g_{t}n_{t}+ \xi\,.
\label{eq:langevin_discrete}
\end{equation}
If $g$ is a damping constant $<1$ and $\xi$ a stationary noise term, this equation is solved by a normally distributed function $\rho(n)$. This is for instance used to derive Boltzmann-Gibbs statistics for particles in an ideal gas. % If $n$ is identified as the momentum of particles $p$, the distribution of the energy $E=\frac{p^2}{2m}$ of free particles decays exponentially.
The Langevin equation has also been widely studied where $g$ is a multiplicative noise term, e.g. by \cite{schenzle1979multiplicative, takayasu1997stable, beck2001dynamical, biro2005power} who give examples from physics and chemistry, where the interpretation of the multiplicative noise are fluctuations of an external field. 
 This explanation has also been applied to the formation of firm size and city size distributions \cite{zanette1997role, marsili1998interacting, gabaix1999zipf}, as well as  to income distribution (e.g. \cite{champernowne1953model,kinsella2011income}). In the presence of a multiplicative noise term $g$ and an additive noise term $\xi$, a variable $n$ whose evolution is described by equation (\ref{eq:langevin_discrete}) exhibits a stationary distribution with a power law tail. $n$ may be a continuous variable or, as in the context of firm growth, discrete, denoting the size of a firm $i$ from an ensemble of firms. The existence of a stationary distribution with a power law decay can be derived from an equation in discrete time (\ref{eq:langevin_discrete}) or continuous time (\ref{eq:langevin_continuous}). In the following, two formalisms taking different approximations are presented.

\paragraph{Derivation via the master equation.}\label{sec:master_equation}
Gabaix \cite{gabaix2008power} shows the existence of power laws for a city size distribution based on an argument by Champernowne in 1953 \cite{champernowne1953model} and developed rigorously by Kesten \cite{kesten1973random}. The counter-cumulative size distribution, which is the probability that a firm $i$ is bigger than a value $x$, is defined as $H_{t+1}(x)=P(n_{i,t+1} >x)$. Its evolution in one timestep, without additive term $\xi$, can be written as
\begin{eqnarray}
H_{t+1}(x)=P(n_{i,t+1} >x)=P(g_{i,t}n_{i,t}>x)=P\left(n_{i,t}>\frac{x}{g_{i,t}}\right)\\
=\int_{0}^{\infty}H_t\left(\frac{x}{g}\right)G(g)d\,g\,,
\end{eqnarray}
where $G(g)$ is the distribution of the growth rate $g$. In a stationary state, $H_{t+1}=H_t$, so the relation becomes
\begin{equation}
 H(n)=\int_{0}^{\infty}H\left(\frac{n}{g}\right)G(g)d\,g\,.
\end{equation}
 For the distribution $H(n)$, \cite{gabaix2008power} and \cite{schenzle1979multiplicative} show that a trial function $H(n)=c/n^\alpha$ ($c=const$) yields the following relation for the noise $g$: $1=\int_{0}^\infty g^{\alpha}G(g)dg$ which is $E[g^{\alpha}]=1$. This holds, however, only if such a stationary state exists, which is only the case if additionally some additive noise $\xi$ is present.
Takayasu et al. \cite{takayasu1997stable} equally derive equation condition $H(\geq n)=c/n^\alpha$,  from the condition of continuity of its characteristic function. 
However, the additive term $\xi$ in (\ref{eq:langevin_discrete}) is needed for the system to achieve a stationary state where  $H(x)$ follows a power-law distribution. For systems of constant global size, where the additive term is small, the noise distribution $\mathcal {G}(g)$ is centered around a value close to 1, so $\left\langle H_{t+1}\right\rangle=\left\langle g\right\rangle\left\langle H_{t}\right\rangle$, so $\left\langle g\right\rangle=1$. For such noise, $\alpha=1$, which is called the Zipf law.

Applied to firm growth, possible interpretations are that the additive term $\xi$ prevents firms from becoming too small, or that some firms are continuously being started and compensate firms that become smaller than a threshold and exit\footnote{If firm size is discrete, this can simply mean that firms which have reached size $n_i=0$ are replaced.}.
Since power law distributions are conserved under addition of a faster decaying distribution \cite{gabaix2008power}, the presence of additive noise does not affect the power law exponent. 

\paragraph{Derivation via the Fokker-Planck equation.}\label{sec:fokker_planck}
The existence of power laws in a stationary state can also be derived by solving the Fokker-Planck equation instead of the master equation \cite{schenzle1979multiplicative, borland1998ito}. Biro and Jakov\'ac \cite{biro2005power}, derive this for the Langevin equation in continuous form:
\begin{equation}
 \dot n + \gamma n= \xi
\label{eq:langevin_continuous}
\end{equation}
with a multiplicative noise term $\gamma$ and additive noise term $\xi$ \footnote{Compared to the discrete equation (\ref{eq:langevin_discrete}), the multiplicative term $g$ corresponds to $1-\gamma$ here.}. Using a method of trial functions introduced by Wang and Uhlenbeck \cite{wang1945theory}, where only terms linear in $dt$ are kept, the Fokker-Planck-equation is derived
\begin{equation}
\frac{\partial f}{\partial t}=-\frac{\partial (F-Gn)}{\partial n}+\frac{\partial^2 ((D-2Bn-Cn^2)f)}{\partial n^2}\,,
\label{eq:fokker_planck}
\end{equation}
where $F$, $G$, $D$, $B$, $C$ are constants standing for mean, variance and cross-correlation of additive and multiplicative noise.

Other approaches leading to the Fokker-Planck equation are \cite{richmond2001power} and \cite{sornette1997convergent}, \cite{marsili1998dynamical}. 
In the limit where the variance of the multiplicative noise goes to zero, but additive noise is present, the stationary solution of (\ref{eq:fokker_planck}) for $n$ is a Gaussian distribution. In the limit where multiplicative noise is present and additive noise goes to zero, the stationary distribution for $1/n$ is a Gamma distribution, which has a power law tail.

\section{Theoretical Analysis}\label{sec:theo_analysis_model}
\subsection{Additive and multiplicative noise}
Additive and multiplicative noise terms in the Langevin equation lead to different stationary size distributions, as results of the corresponding Fokker-Planck equation show in two limits (\ref{sec:fokker_planck} and \ref{sec:implementation_noise} for illustrations). If $f(n)$ is thought of as the firm size distribution, additive fluctuations tell the absolute change in size of a firm,  and multiplicative fluctuations tell the ratio by which a firm's size has changed, i.e. they are relative fluctuations.  These two have different $n$-dependencies, which are both different to the $n$-dependency of the model introduced in section \ref{sec:model}. $n$-dependency of the standard deviation of absolute and relative fluctuations are compared for additive noise, multiplicative noise and the introduced model in table \ref{tab:add_mult}.
\begin{table}[h]
\label{tab:add_mult}
\small
 \renewcommand{\arraystretch}{1.5} 
\begin{tabular}{llll}
\toprule

\vspace{-0.2cm}&$\sigma$ of &$\sigma$ of &stationary\\
&relative fluctuations&absolute fluctuations&size distribution\\
\hline
\vspace{-0.2cm}
purely&&\\
additive noise&$\propto 1/n$&$const$&Gaussian\\
\vspace{-0.2cm}
purely&&\\
mutliplicative noise&$const$&$\propto n$&Power law\\
studied model&$\propto 1/\sqrt{n}$&$\propto\sqrt{n}$& Fat-tailed \\
\toprule
\end{tabular}
\caption{Different $n$-dependencies for additive and multiplicative noise in comparison to the one from the introduced model.}
\label{tab:add_mult}
\end{table}

For the introduced model, the time evolution of the firm size distribution cannot be described by equation (\ref{eq:langevin_discrete}), where $g$ and $\xi$ are drawn from distributions independent of $n$. Here, the standard deviation $\sigma$ of the noise depends on $n$. It may either be written as additive Gaussian noise with $\sigma_{add}\propto \sqrt{n}$
\begin{equation}
 n_{t+1}=n_t+\xi(\sqrt{n})\,,
\end{equation}
where $\xi$ is Gaussian with $\sigma\propto \sqrt{n}$, or as multiplicative Gaussian noise with $\sigma_{mult}\propto \frac{1}{\sqrt{n}}$, 
\begin{equation}
 n_{t+1}=g(\frac{1}{\sqrt{n}})n_t
\label{eq:formulation_multiplicative}
\end{equation}
where $g$ is Gaussian with $\sigma\propto\frac{1}{\sqrt{n}} $ noise. %Table \ref{tab:add_mult} shows this $n$-dependence of the noise in the model in comparison to purely additive and multiplicative noise. 
Formulation (\ref{eq:formulation_multiplicative}) has been chosen for the equations (\ref{eq:gaussian_growthrate_independent}) and (\ref{eq:gaussian_growthrate_independent_goods}). It corresponds to a more complicated stochastic process that has unfortunately not yet been solved analytically. Also in the literature, only few solutions for more complicated processes exist \cite{schenzle1979multiplicative}. A different process whose noise has the same $n$-dependency as this model is treated in \cite{marsili1998interacting}, and discussed further in section \ref{sec:discussion}. %In the following section, some numerical results are presented and discussed.

\subsection{The growth rate probability distribution}\label{sec:growth_rate_theoretical}
Figure \ref{fig:78aBIGgrowth} shows that despite the normally distributed $\mathcal{G}(g|n)$, the aggregate $\mathcal{G}(g)$ exhibits a tent-shape. The $n$-dependence (equation \ref{eq:scaling_relation}) is the reason why the growth rate distribution of $\mathcal{G}(g|n)$ is wider for small firms and more narrow for big firms. All firms contribute to the aggregate $\mathcal{G}(g)$ 
 and the growth rate distribution for the $N_f$ firms can be written as 
\begin{equation}
 \mathcal{G}(g)=\frac{1}{N_f}\sum_{i=1}^{N_f}\mathcal{G}(g_i | n_i)\,,
\end{equation}
 or,  in the continuous limit:
\begin{equation}
  {\mathcal{G}}(g)=\int_{0}^{\infty}d n \mathcal{G}(g|n)\rho(n)\,,
\label{eq:growth_rate_integral}
\end{equation}
where $\rho(n)$ is the firms' size distribution. Since so far no analytical expression for the simulated size distribution of this model is derived, we evaluate the integral as an approximation for power-law size distributions of exponent $\alpha$. For firms' size distributions $\rho(n)\propto n^{-\alpha-1}$ and scaling exponents $\beta =0.5$
the integral yields
\begin{equation}
  \mathcal{G}(g)\propto \int_{0}^{\infty}n^{0.5}\frac{1}{n^{\alpha+1}}\frac{1}{\sqrt{2\pi}}e^{-\frac{1}{2}n\,(g-1)^2}dn=\frac{1}{\sqrt{\pi}}2^{-\alpha}\,(g-1)^{2\alpha -1}\Gamma\small{\left(\frac{1}{2}-\alpha\right)}\,.
\label{eq:growth_rate_evaluated}
\end{equation}

For $\alpha=0$, it simplifies to $ \mathcal{G}(g) = \frac{1}{|g-1|}$. In this case the slopes of $\mathcal{G}(g)$ are linear on a double-logarithmic scale, i.e. presenting a tent-shape. For $\alpha >0.5$, the integral diverges at the origin and can only be integrated starting from a finite cutoff value $n_0$, since $n^{-\alpha-1}$ is not a normalized probability density. This includes the case of our model (see section \ref{sec:num_results_all}, $\alpha\approx 0.7$ and $\beta$ = $0.5$). If integrated from a cutoff $n_0$, equation (\ref{eq:growth_rate_evaluated}) still yields a tent-shaped $\mathcal{G}(g)$, with its width depending on the cutoff. Integral (\ref{eq:growth_rate_evaluated}) can be generalized to values of $\beta $ other than 0.5, which is interesting since empirical values are $\alpha\approx 1$ and $\beta$ $\approx 0.25$. In this case, the condition for convergence of the integral becomes $\alpha>\beta $ , and for $\alpha=0 $ the expression simplifies to $ \mathcal{G}(g) = \frac{1}{2\,\beta \,|g-1|}$. The smaller is $\beta $ , the less peaked $\mathcal{G}(g)$, which is intuitive, since if $\beta$ = $0$, the result is a Gaussian $\mathcal{G}(g)$. 

Notice that the shape of $\mathcal{G}(g)$ is not very sensitive to the underlying size distribution: equation (\ref{eq:growth_rate_evaluated}) yields an approximate tent-shaped $\mathcal{G}(g)$ even for exponential decay of $\rho(n)$. This suggests that despite the fact that the size distribution in the proposed model deviates from a Zipf law, the idea of performing integral (\ref{eq:growth_rate_integral}) explains the observed tent-shape well.

In the literature, the principle of performing this integral has been used in the model by \cite{fu2005growth} to obtain a tent-shaped growth rate distribution of a single firm. Other models \cite{stanley1996scaling, lee1998universal, bottazzi2006explaining, picoli2006scaling} do not perform the integral and do not clearly distinguish between the growth rate probability at firm level and at aggregate level. Both are fitted with a Laplacian distribution. The necessity to perform the integral in equation (\ref{eq:growth_rate_integral}) is however independent of the assumed $\mathcal{G}(g|n)$. The functional form of $\mathcal{G}(g)$ yields an approximate $1/|g-1|$ tent-shape for both Laplacian and Gaussian $\mathcal{G}(g|n)$. If it is integrated from a size cutoff, for low values of $\beta $, Laplacian $\mathcal{G}(g|n)$ yield a more peaked aggregate $\mathcal{G}(g)$.  Since the growth rates typically have values close to 1, empirical evidence can often be fitted equally well with a Laplacian (centered around 1) and a $1/|g-1|$-function. However \cite{fu2005growth} find that the tails of the tent-shape exhibit power law decay rather than exponential decay of a Laplacian, substantiating the argument presented here.

\section{Numerical Results}\label{sec:num_results_all} Technical issues are detailed in appendix \ref{sec:implementation}.
\subsection{Results I -- Size distribution}\label{sec:num_results1}
\begin{figure}[h!]
	\centering
\begin{subfigure}[b]{0.45\textwidth}
\includegraphics[angle=270, width =\textwidth]{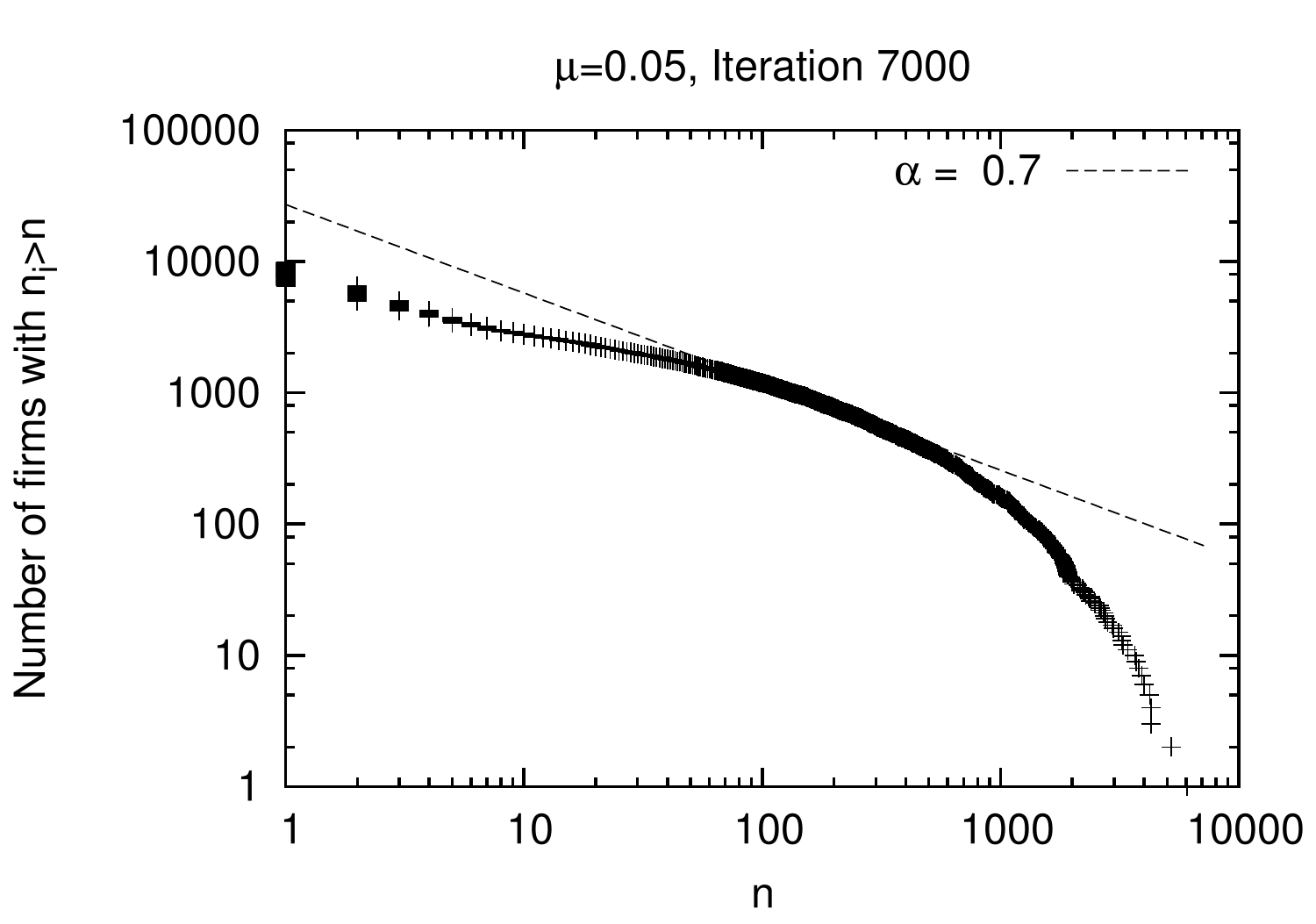}%\hspace{0.5cm}
\caption{}
\end{subfigure}
\begin{subfigure}[b]{0.45\textwidth}
\includegraphics[angle=270, width =\textwidth]{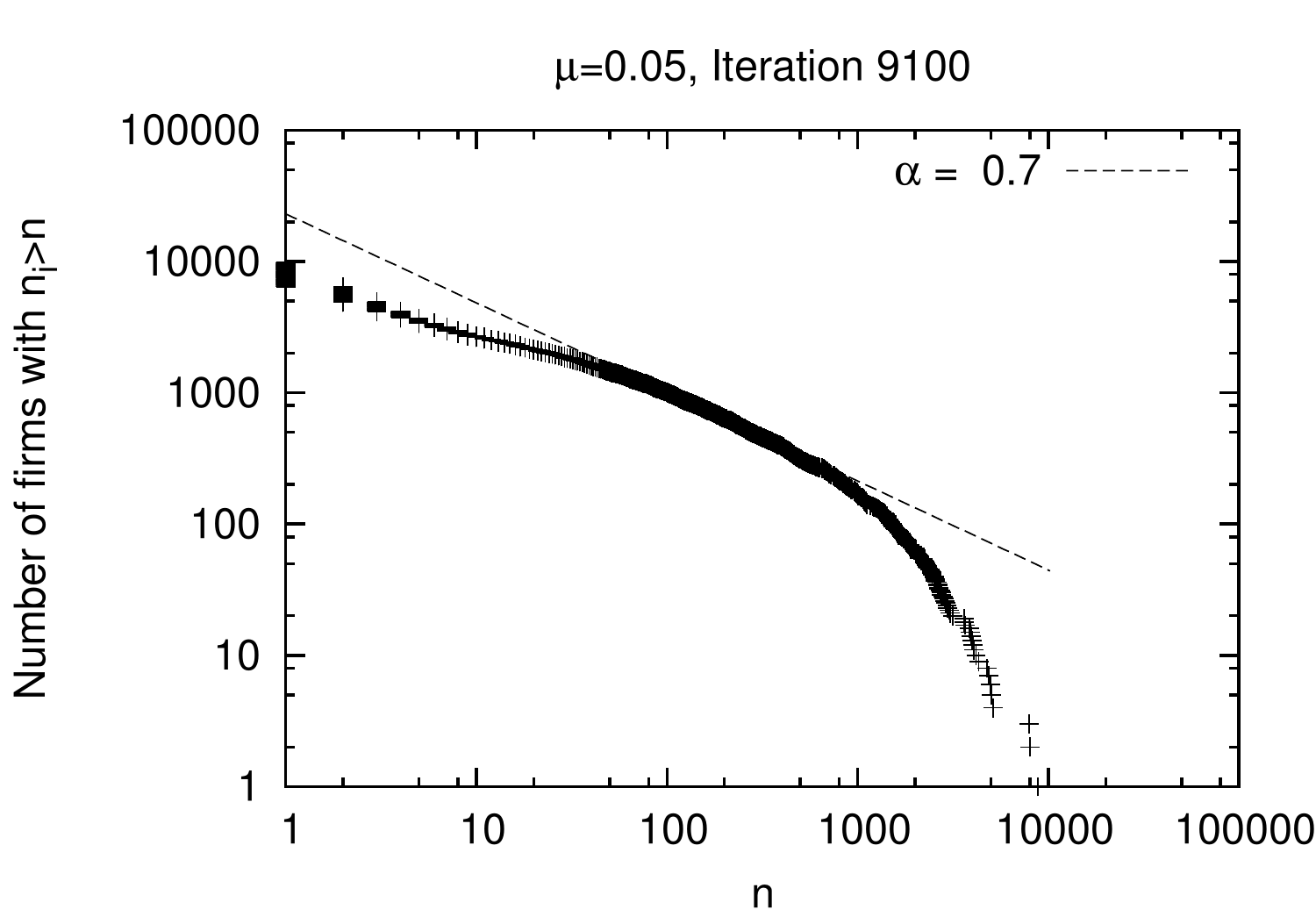}%\hspace{0.5cm}
\caption{}
\end{subfigure}

\begin{subfigure}[b]{0.45\textwidth}
\includegraphics[angle=270, width =\textwidth]{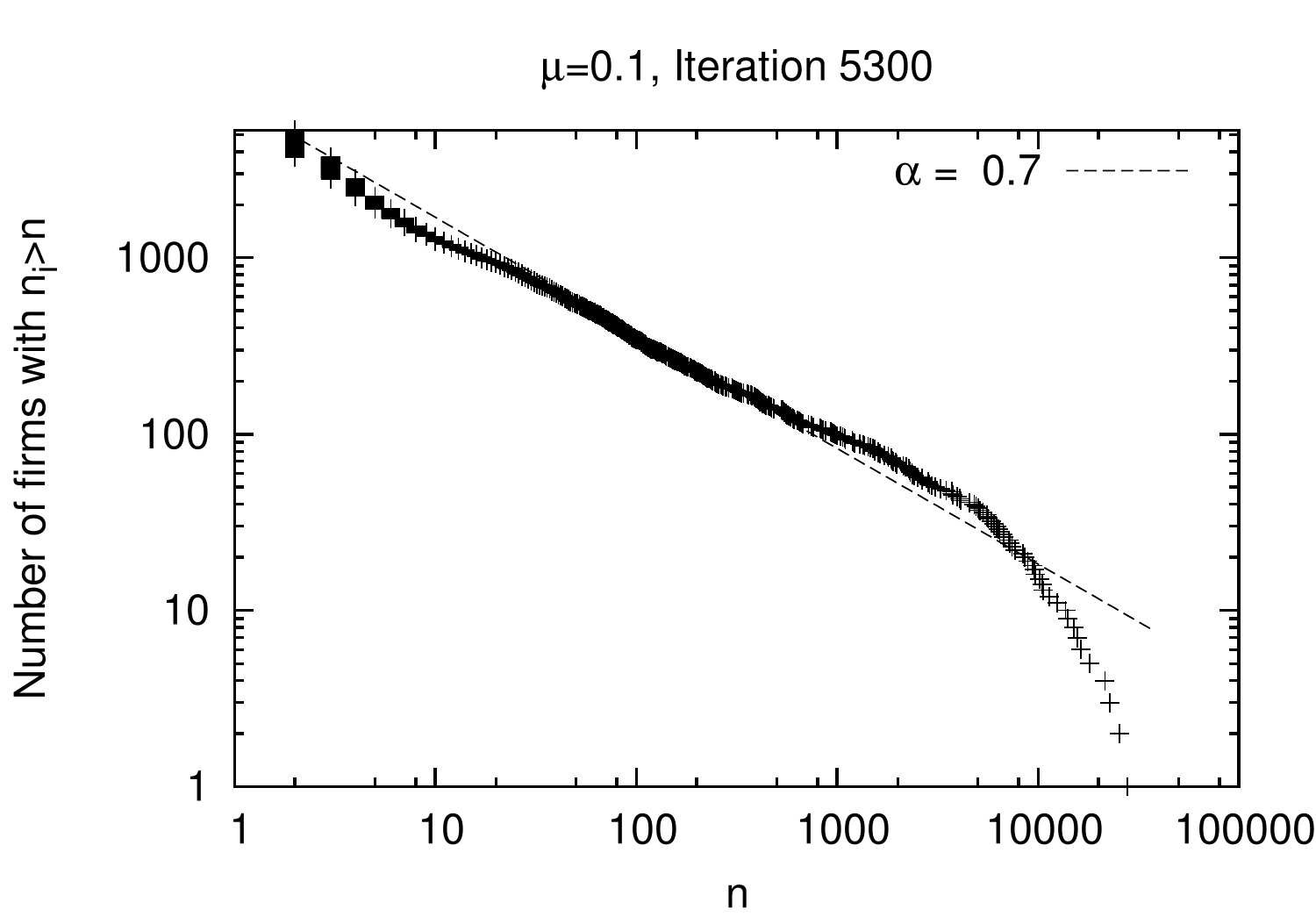}
\caption{}
\end{subfigure}
\begin{subfigure}[b]{0.45\textwidth}
\includegraphics[angle=270, width =\textwidth]{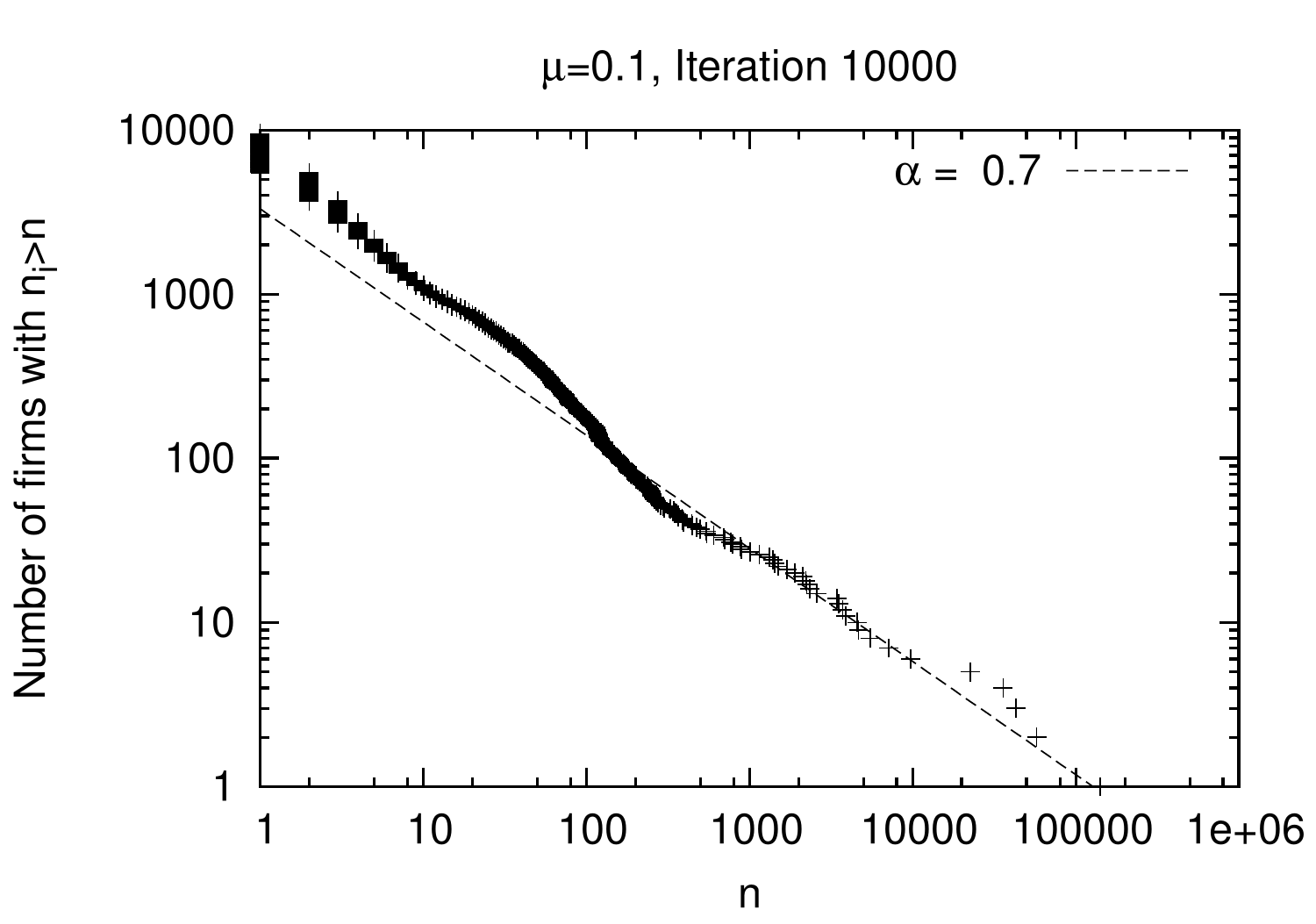}
\caption{}
\end{subfigure}

\begin{subfigure}[b]{0.45\textwidth}

\includegraphics[angle=270, width =\textwidth]{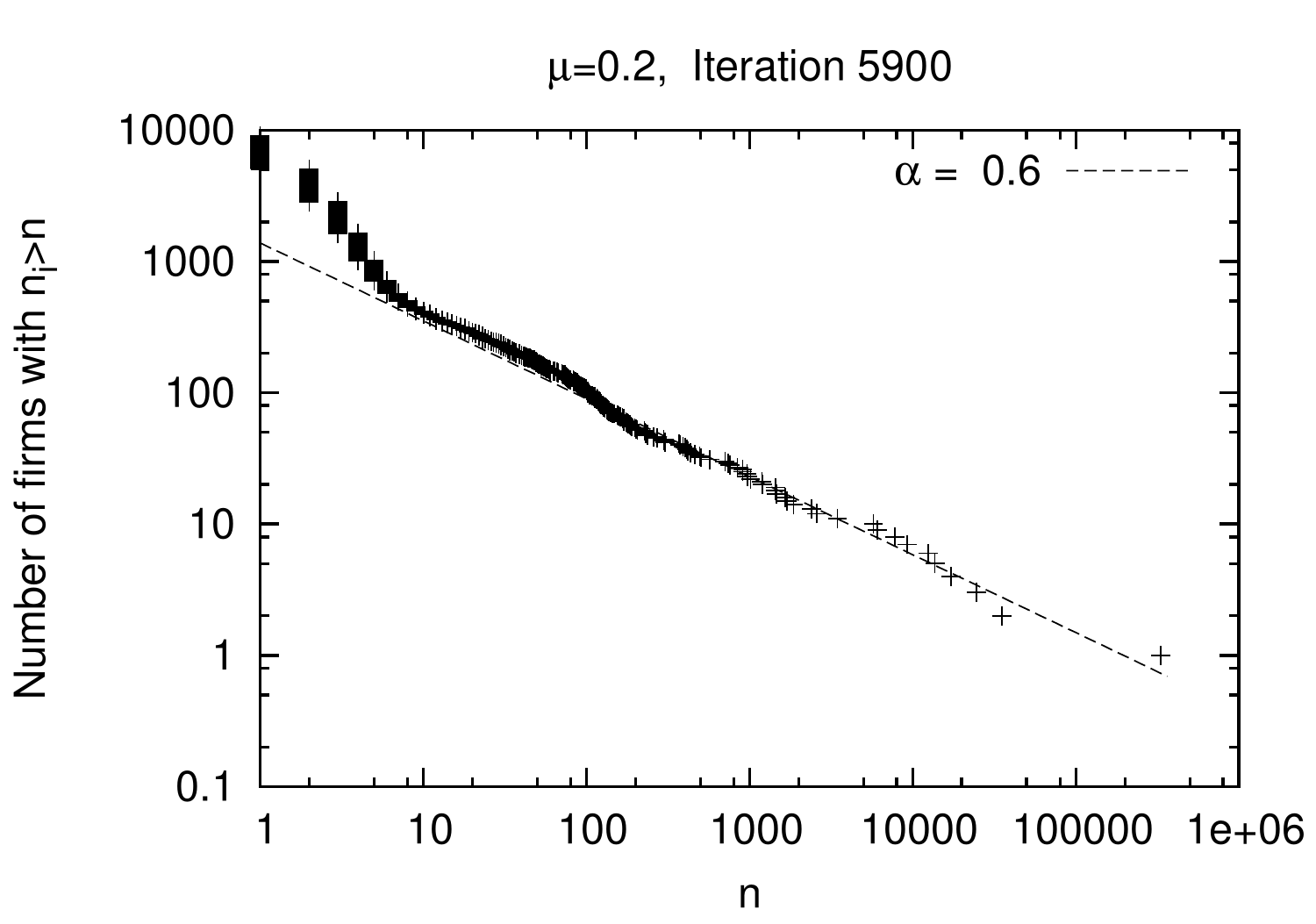}
\caption{}
\end{subfigure}
\begin{subfigure}[b]{0.45\textwidth}
\includegraphics[angle=270, width =\textwidth]{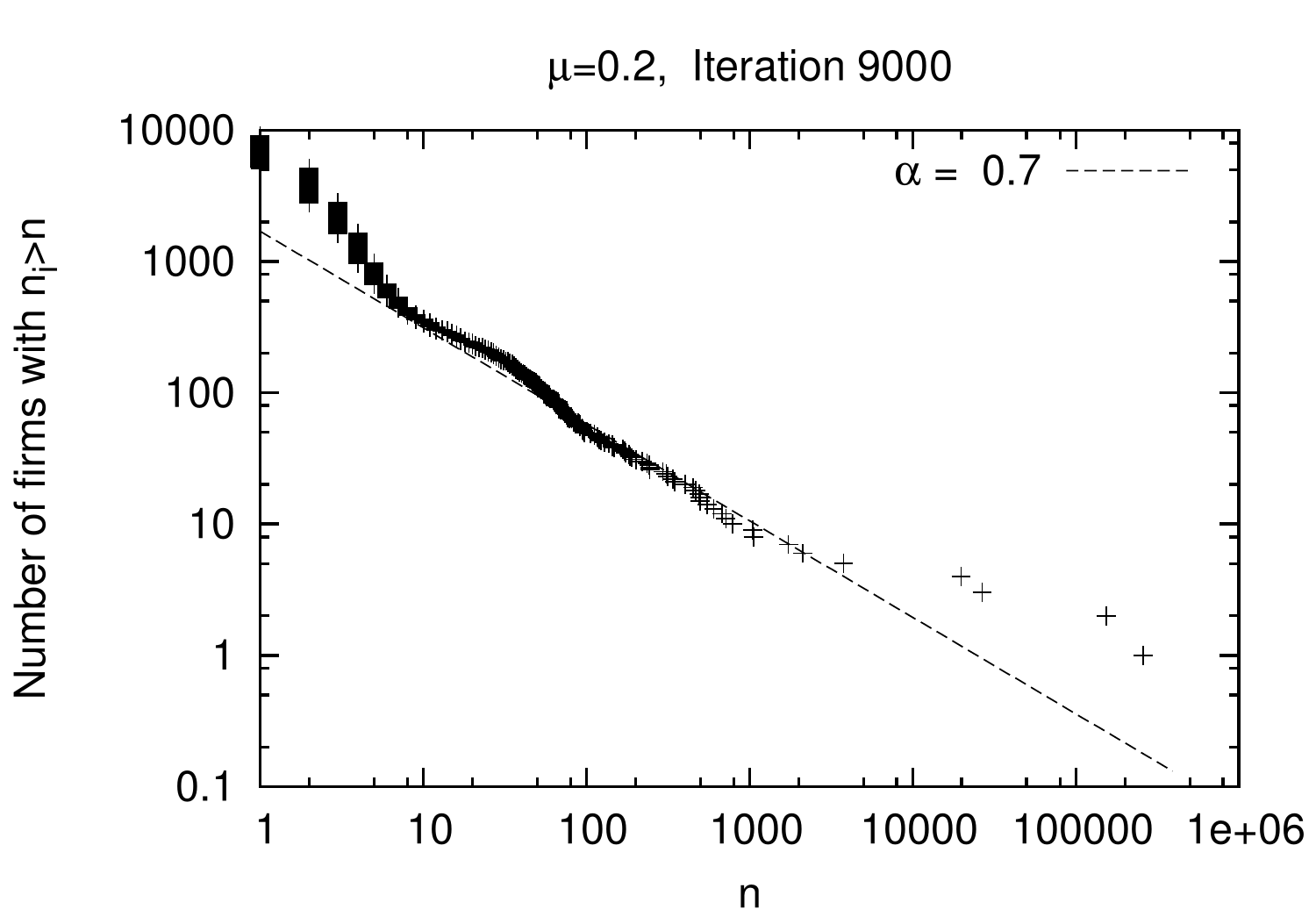}
\caption{}
\end{subfigure}
\begin{subfigure}[b]{0.45\textwidth}
\includegraphics[angle=270, width =\textwidth]{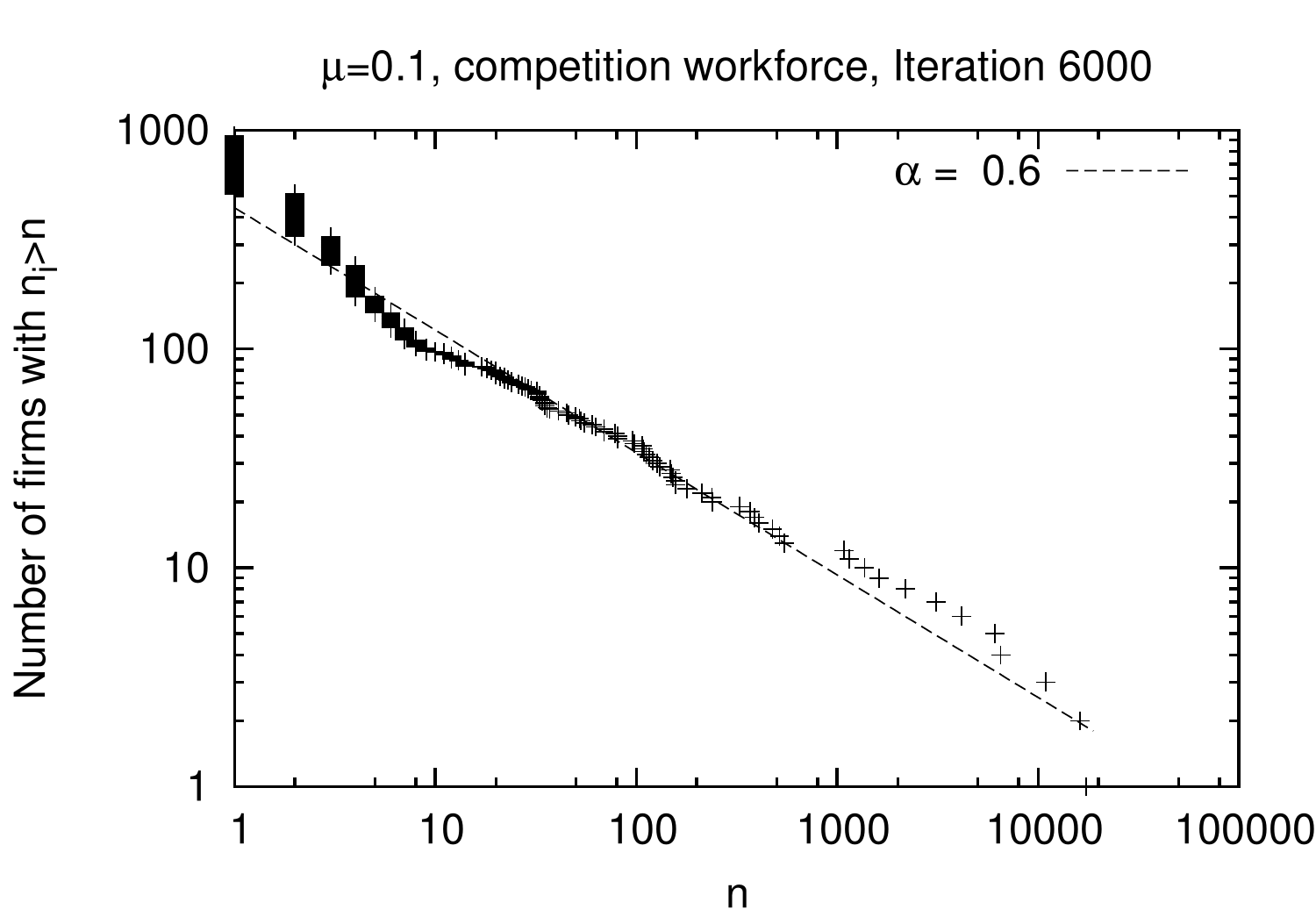}
\caption{}
\end{subfigure}
\begin{subfigure}[b]{0.45\textwidth}
\includegraphics[angle=270, width =\textwidth]{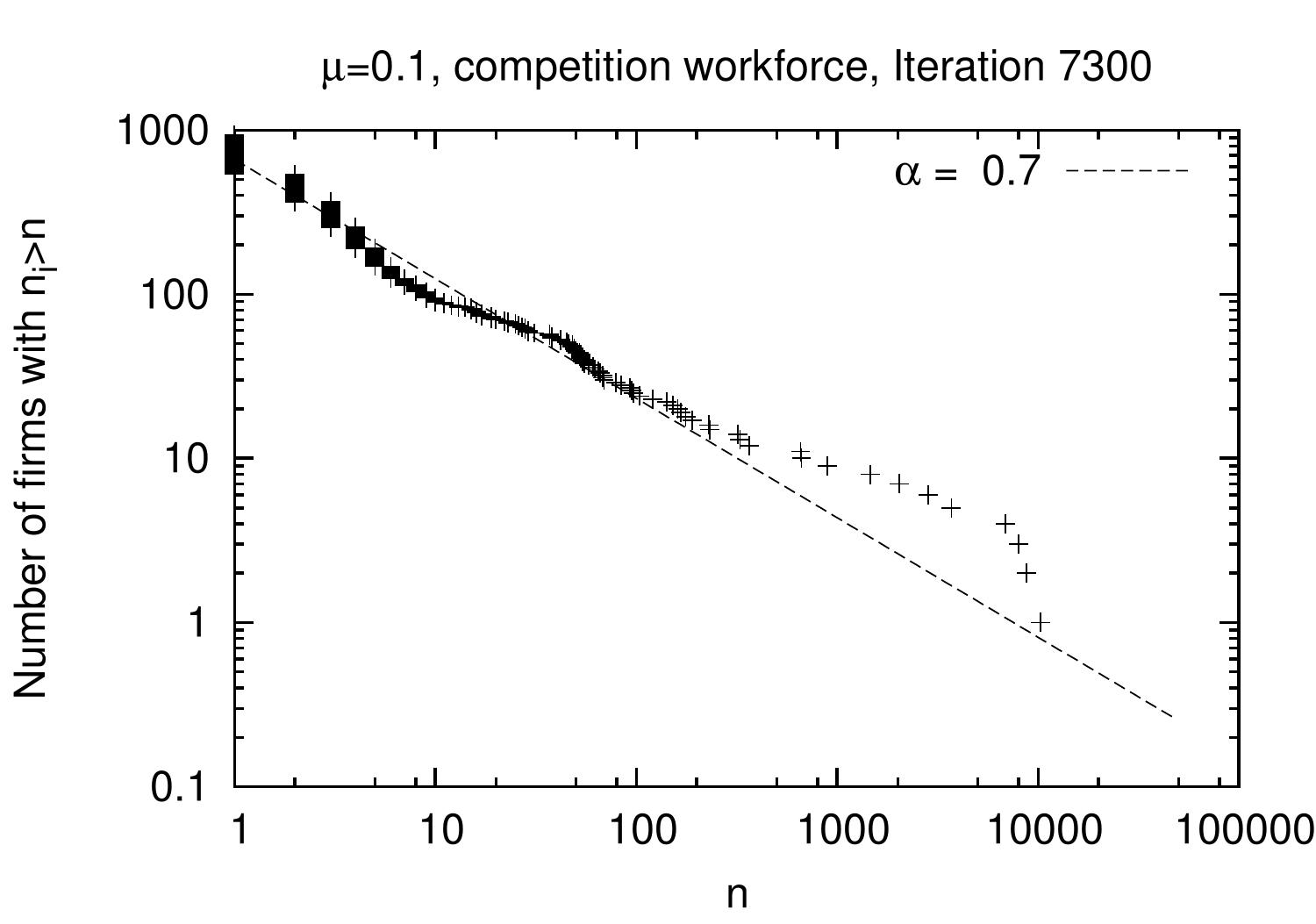}
\caption{}
\end{subfigure}
	\caption{Four examples of the evolution of firm size distribution, at different steps of convergence. The first three rows are scenario (ii) where firms do not spend, the fourth row (i) where firms spend. First row: $\mu=0.05$, second row: $\mu=0.1$, third row: $\mu=0.2$, fourth row: $\mu=0.1$, firms spend their profits. The power law exponent is $<1$. Extinct firms are replaced by one of size $n\in [1,2]$ with the method detailed in \ref{sec:implementation_noise}. For a value of $\mu=0.05$, the size distribution can be fitted with a power law of exponent $\approx 0.7$, too, but the tail stays concave and does not converge to it (see subfigures (a) and (b)). }
	\label{fig:rank_size_examples}
\end{figure}

In figure \ref{fig:rank_size_examples}, typical examples of the time evolution of a scenario where the only constraint is limited purchasing power in the goods market are shown. To keep the system at constant global size, firms that reached size $0$ were re-started $n\in[1,2]$ ($1.5$ on average), by the method detailed in \ref{sec:implementation_noise}.
Compared to simulations of systems with multiplicative noise, simulations of this model take much longer to approach a power law, whose exponent is $\alpha\approx 0.7$, which is much lower than the exponent $1$ of a Zipf law. $\alpha=1$ is the lowest exponent that is found with multiplicative noise (see figure \ref{fig:pure_multiplicative_noise}). This flatter power law decay has been found for different values of $\mu$ in the range $0.05 \leq \mu\leq 0.2$, as well as for both scenarions (i) and (ii). As shown in successive snapshots in figure \ref{fig:rank_size_examples}, the size distribution fluctuates much more than for purely multiplicative noise shown in \ref{fig:pure_multiplicative_noise} in appendix \ref{sec:implementation_noise}. 

\begin{figure}[h!]
	\centering
\begin{subfigure}[b]{0.9\textwidth}
\includegraphics[angle=270, width =\textwidth]{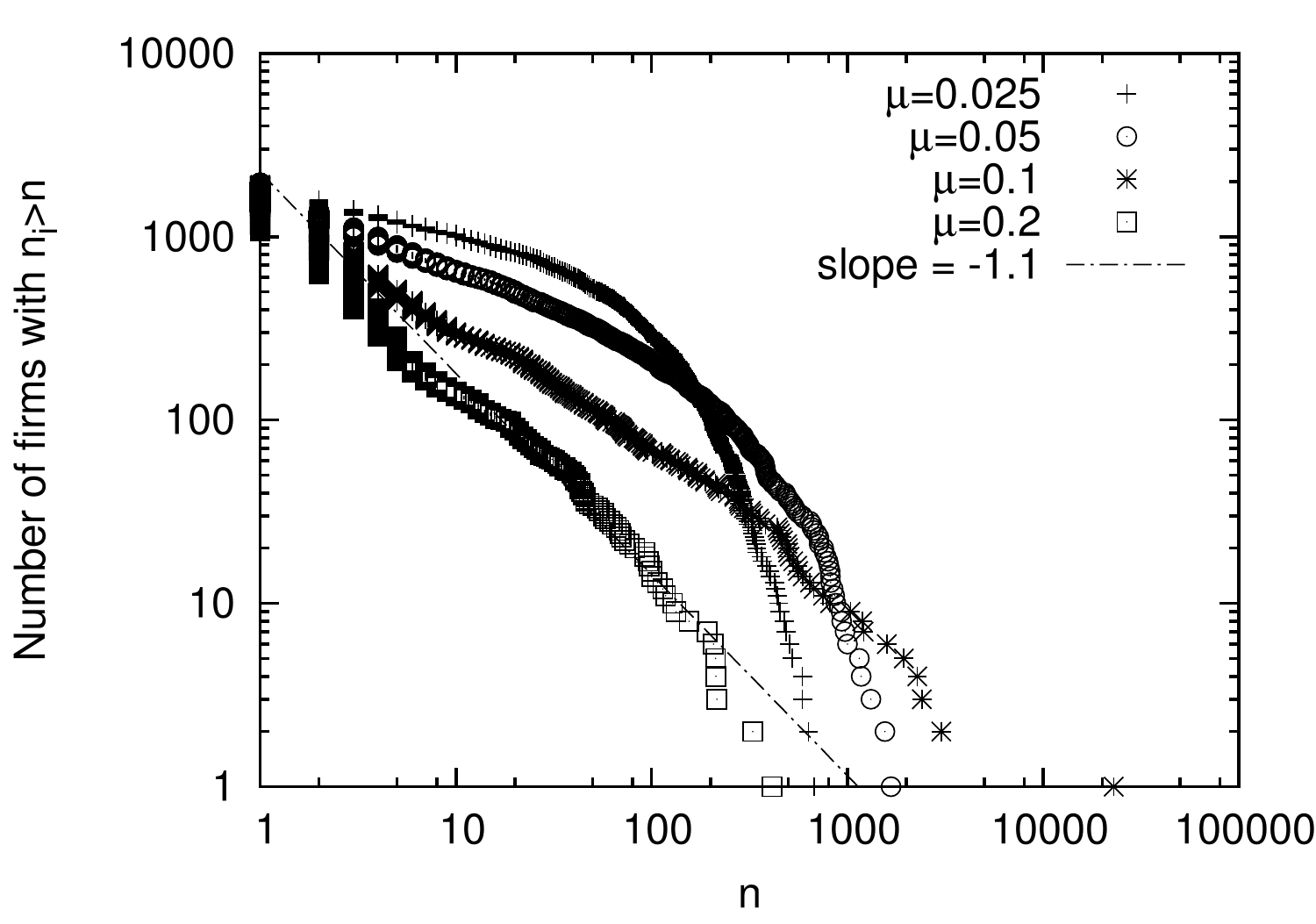}
\caption{}
\end{subfigure}

	\caption{Firm size distribution after $5000$ iterations for a system of $2\cdot 10^3$ firms and $9\cdot 10^4$ workers for different values of $\mu$. 
Firms do not consume (scenario ii detailed in section \ref{sec:goods_market}); growth rates are binomial (the method in section \ref{sec:rounding_joboffer} is not used).}
	\label{fig:rank_size_n1_n50}
\end{figure}
For small systems (e.g. $2000$ firms), the size distribution does not converge to a power law for low values of $\mu$, but has been found to stabilize with a faster decay than a power law. The interpretation is that for small systems, as well as for low $\mu$, the rounding introduced by equation (\ref{eq:probabilistic_rounding}) modifies the actual planned production $\hat q_i$ more than the multiplication by $(1+\mu^{net}_{i, t-1})$. The modification through rounding can be regarded as additive noise rather than the noise with $\sigma\propto n^{-1/2}$- scaling. %
This effect is stronger for smaller systems.
\clearpage
\subsection{Results II -- Growth rate probability distribution}\label{sec:growth_rate}
\begin{figure}[h!]
	\centering
\begin{subfigure}[b]{0.6\textwidth}
\includegraphics[angle=270, width =\textwidth]{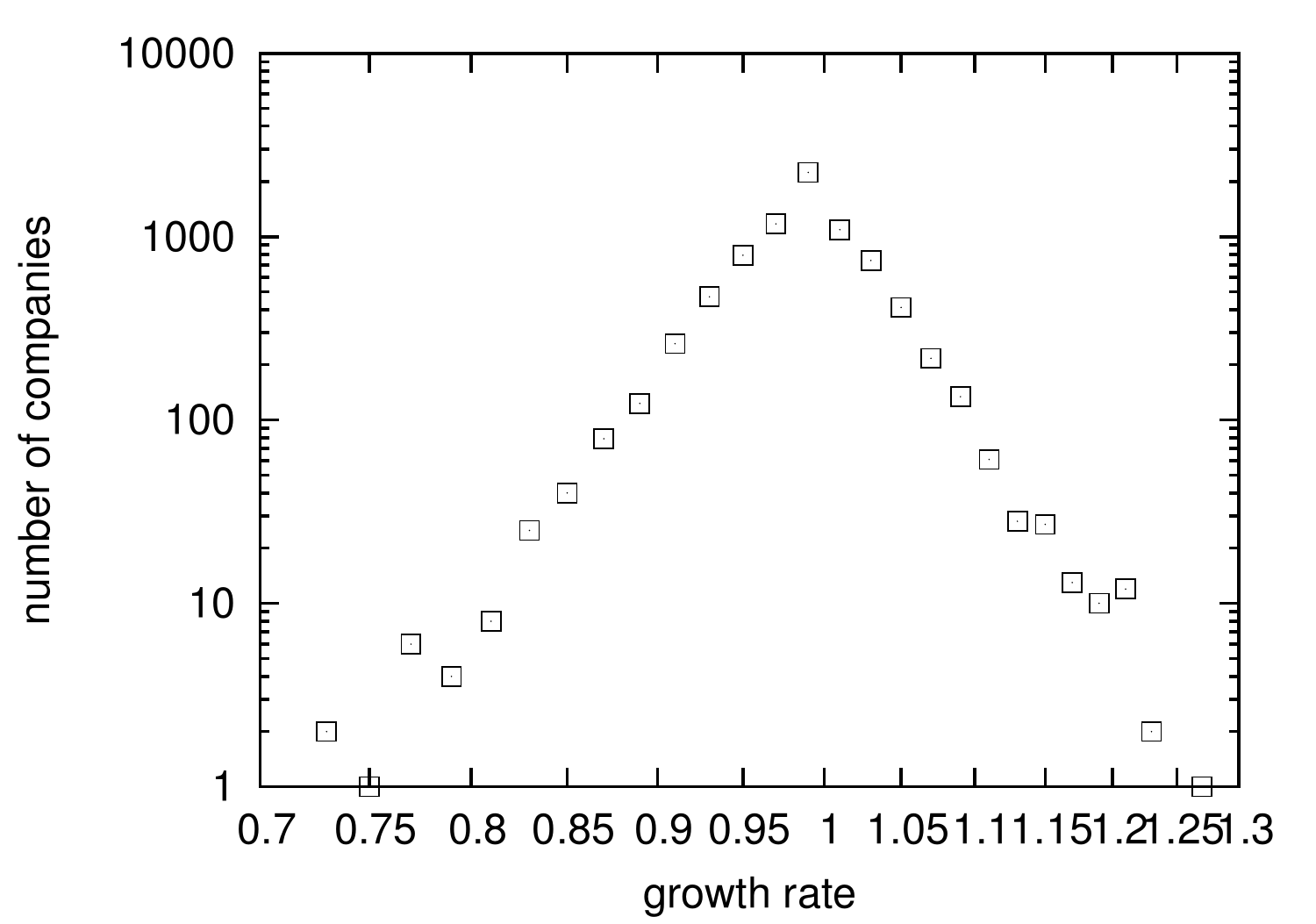}
\caption{}
\end{subfigure}

	\caption{(a) growth rate distribution with rounding detailed in appendix \ref{sec:rounding_joboffer}, in a system with full employment, where the only constraint is workforce availability. (b) corresponding size distribution. $10^4$ firms, $10^6$ workers.}
	\label{fig:78aBIGgrowth}
\end{figure}
Figure \ref{fig:78aBIGgrowth} shows the tent-shaped growth rate distribution. Its explanation, detailed in \ref{sec:growth_rate_theoretical} uses a superposition of Gaussian $\mathcal{G}(g|b)$ of $n$-dependent standard deviation. Small firms have larger $\sigma$, which is why they account for the `fat tails', whereas big firms dominate in the peak of the growth rate distribution.

\paragraph{Artefacts from binning.}
Some effects arising from binning data are addressed here. Empirical data in \cite{stanley1996scaling, lee1998universal, sutton2002variance, picoli2006scaling} exhibits tent-shaped growth rate distributions of different widths depending on firms size (or country's size or citations respectively). For all of these, a Laplacian fit is proposed. To do this, firms are grouped according to their size in large logarithmic bins. From the slopes of the growth rate distribution on logarithmic scale, $\sigma (n)$ and its scaling exponent $\beta $ are determined.

\begin{figure}[h!]
	\centering
\begin{subfigure}[b]{0.45\textwidth}
\includegraphics[angle=270, width =\textwidth]{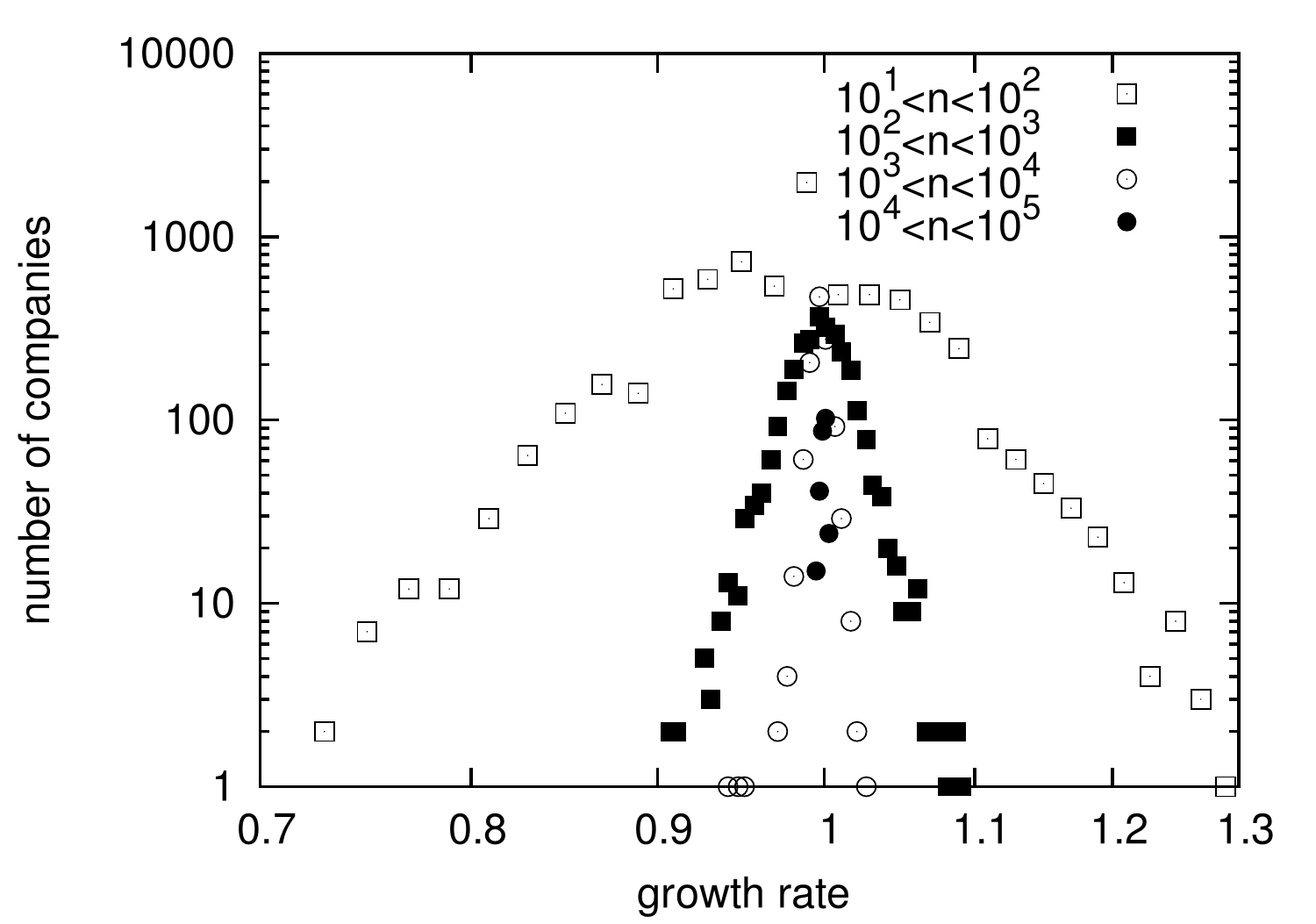}%\label{52BIGfig2}
\caption{}
\end{subfigure}
\begin{subfigure}[b]{0.45\textwidth}
\includegraphics[angle=270, width =\textwidth]{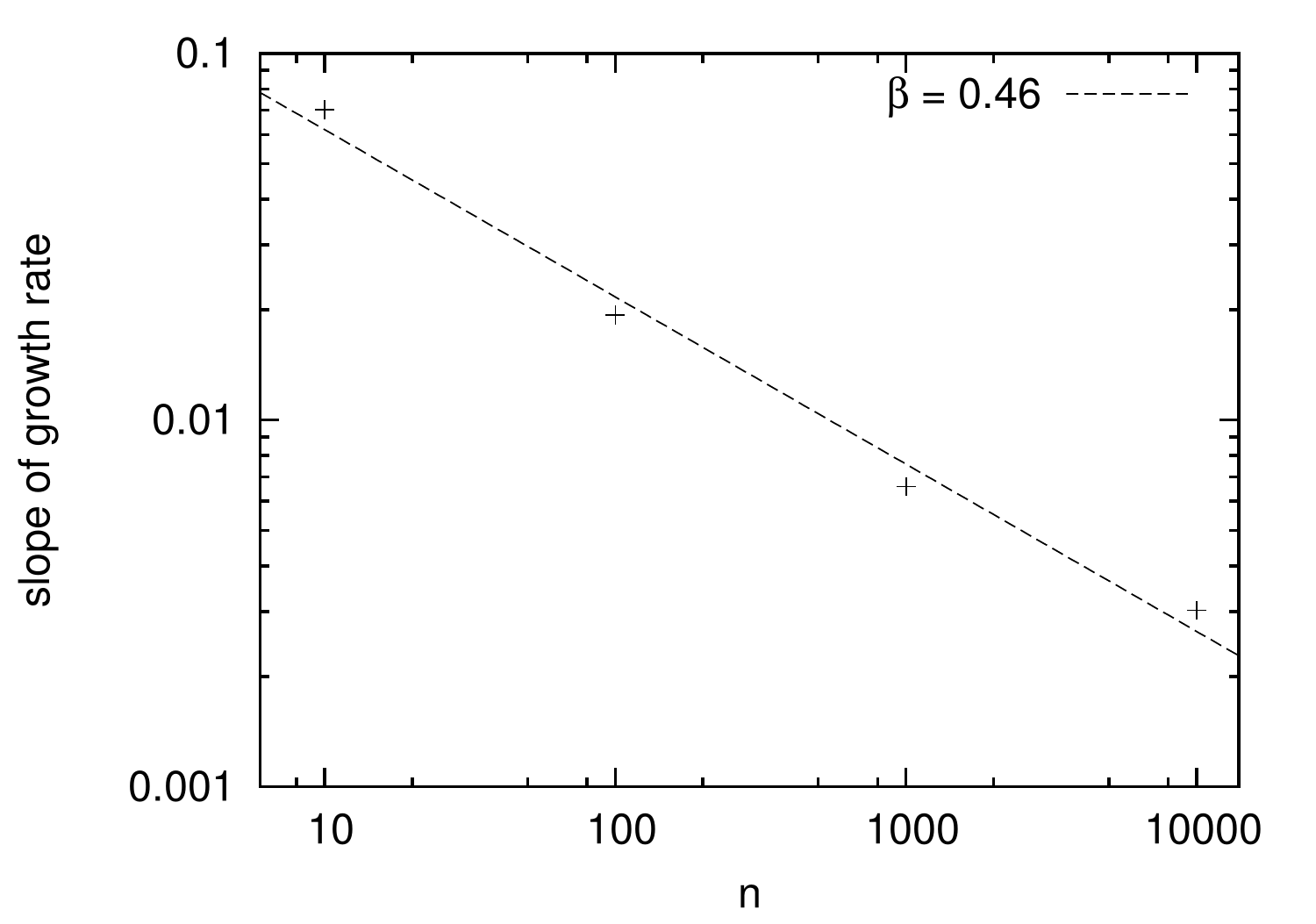}
\caption{}
\end{subfigure}
 	\caption{Histograms for growth rate distributions. (a) Aggregate growth rate distribution for companies ranging from 1 to $5\cdot10^4$ employees from a system with $10^7$ employees and a Zipf firm size distribution. (b) For Gaussian growth rate probability densities, the growth rate distribution of firms within one order of magnitude (its smallest size indicated in the plot) resembles a tent-shape, thus in qualitative agreement with the data shown in \cite{stanley1996scaling}. (c) The slopes of these approximate $1/|g-1|$-distributions follow the same scaling relation as the variance of the Gaussian growth rate probability densities $\sigma(n)\propto n^{-0.5}$. }
	\label{fig:67}
\end{figure}

 Numerical simulations of the model show that aggregation of firm growth rates within one order of magnitude of size is sufficient to obtain a growth rate distribution that resembles a tent-shape, when $\mathcal{G}(g|n)$ is Gaussian (see figure \ref{fig:67}). The reason for this is that $\rho(n)\approx 1/n^{1.7}$  and $\sigma(n)\propto n^{-0.5}$. This result implies that if the average of an ensemble of firms is used to determine the shape of $\mathcal{G}(g|n)$, its functional form is only assessed correctly if the sampled firms have precisely the same size. If $\mathcal{G}(g)$ is $n$-dependent, a size spectrum of one order of magnitude is already enough to modify the form of $\mathcal{G}(g|n)$. The value of $\beta $ does not seem to depend on binning. By plotting the slope of the obtained $\mathcal{G}(g|n_{max})$ against $n_{max}$ of the respective bin, the found relation is again $\sigma \propto n^{-0.5}$.

The very simple underlying microscopic mechanisms suggest that Gaussian functions might be a simpler alternative to the commonly assumed Laplacian shape for $\mathcal{G}(g|n)$, since it also yields a tent-shaped $\mathcal{G}(g)$. Furthermore, Gaussian distributions are conjugate priors to themselves, so may result from several reasons, where each is Gaussian distributed.

\section{Results with different values of $\beta $}\label{sec:scaling_relation}
The empirical studies cited in the introduction find smaller values for $\beta $ than $0.5$, the value in this model. These are often explained by firm-intern factors contributing to a firm's growth, as by \cite{stanley1996scaling,nunes1997scaling, wyart2003statistical, sutton2002variance}. 
Intuitively, if the growth of a company was entirely dependent on the decisions of its CEO, there would be no reason to assume that a company's size should affect its growth rate variance. Under this assumption, values of $\beta $ between $0$ and $0.5$ are possibly due to a contribution of both internal hierarchical structure and of a firm's size. 
For comparison to these empirical results, we simulated a system with a Gaussian $\mathcal{G}(g|n)$ and a scaling exponent $\beta =0.25$. This $\beta $ is not the result of interactions in the job market, but firm's growth consists Gaussian multiplicative noise where $\sigma \propto n^{-0.25}$, without specifying its microfoundations. 
\begin{figure}[h!]
	\centering
\begin{subfigure} [b]{0.45\textwidth}\includegraphics[angle=270, width =\textwidth]{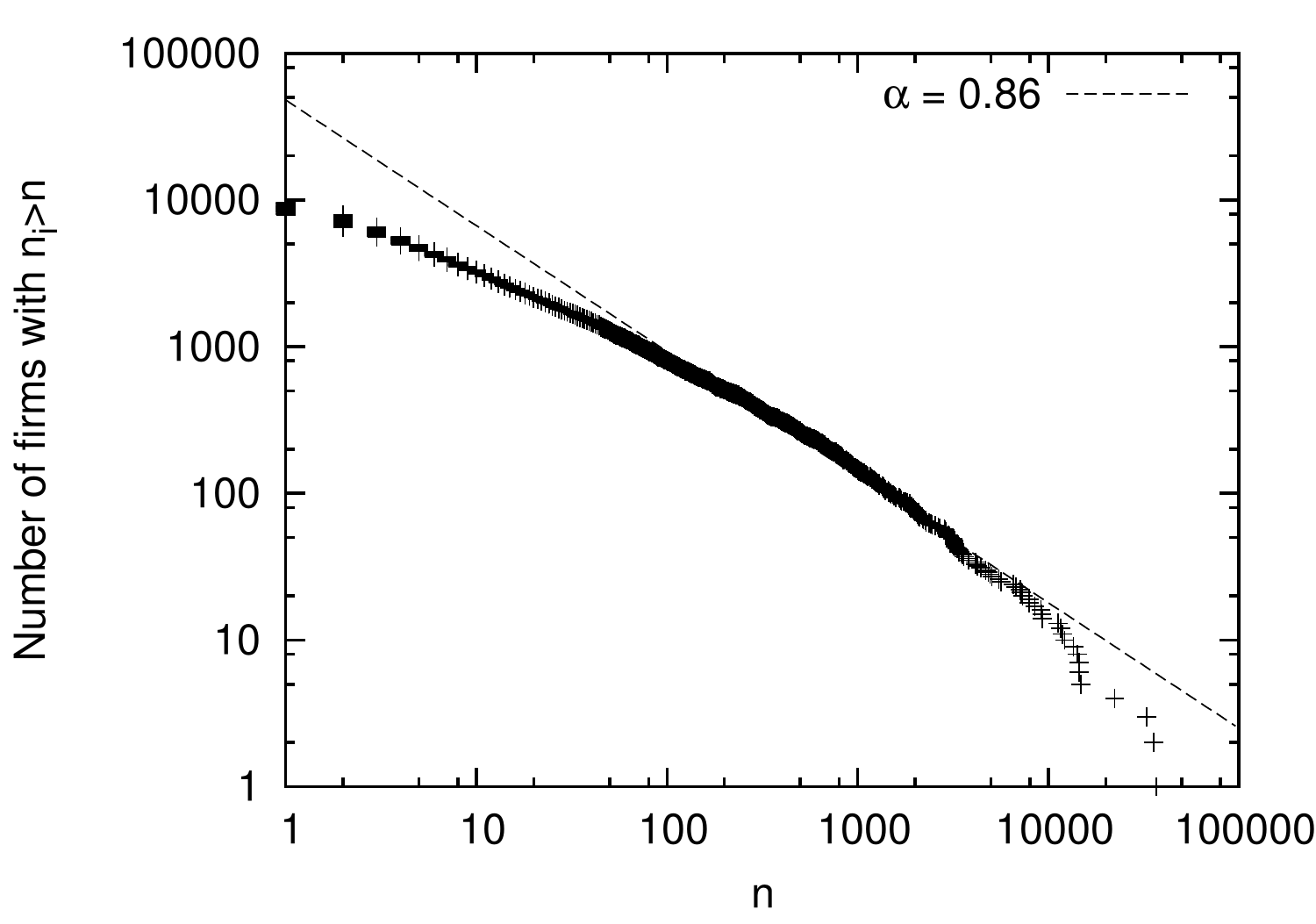}\caption{}\end{subfigure}
\begin{subfigure} [b]{0.45\textwidth}\includegraphics[angle=270, width =\textwidth]{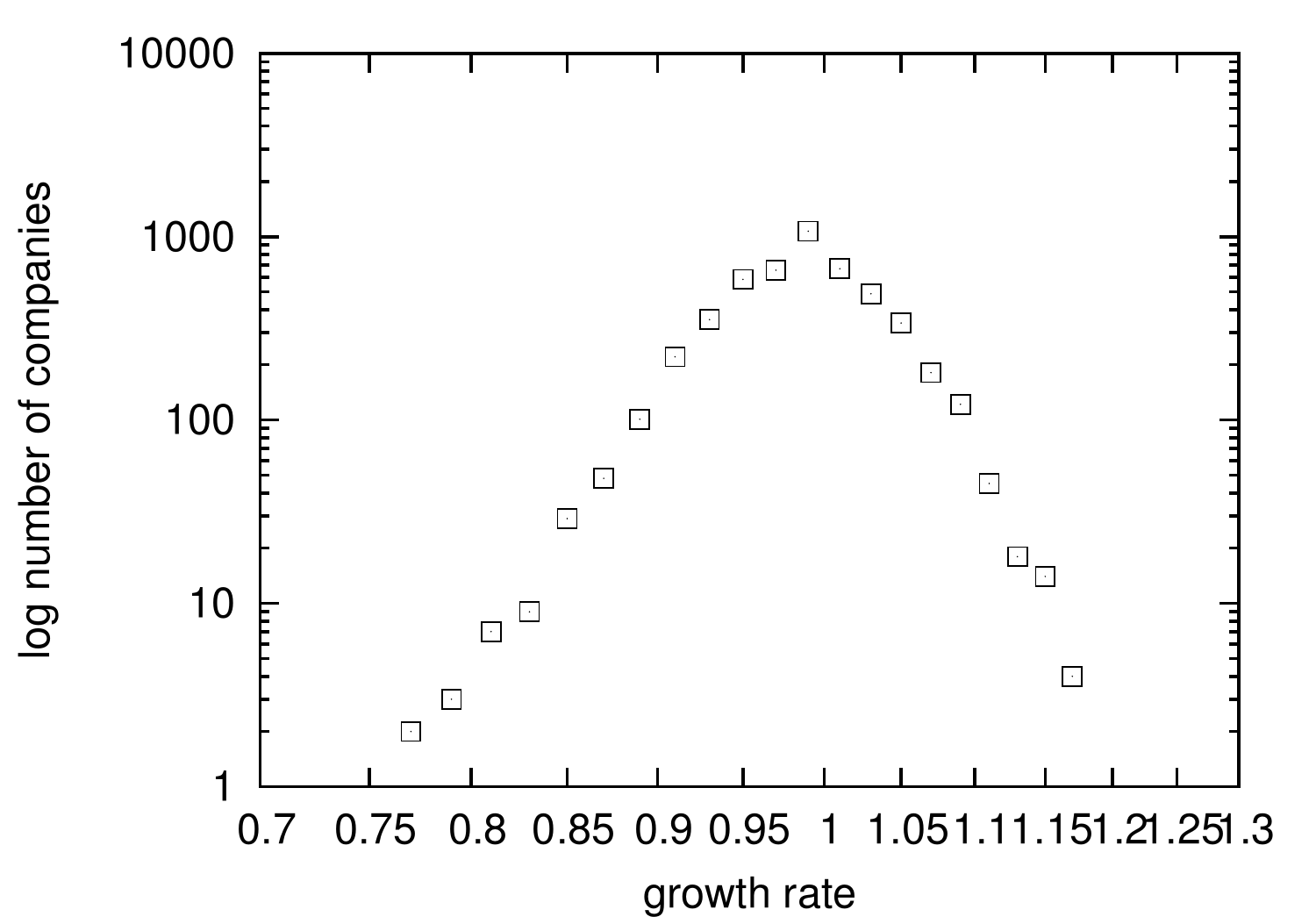}\caption{}\end{subfigure}
 	\caption{ (a) counter-cumulative size distribution and (b) growth rate distribution of a simulation with Gaussian multiplicative noise with a scaling relation of $\beta $ = $0.25$.  (after $3000$ iterations in a system with $10^6$ workers and $10^4$ companies)}
	\label{fig:85cBIGsim}
\end{figure}

\begin{figure}[h!]
	\centering
\begin{subfigure} [b]{0.45\textwidth}\includegraphics[angle=270, width =\textwidth]{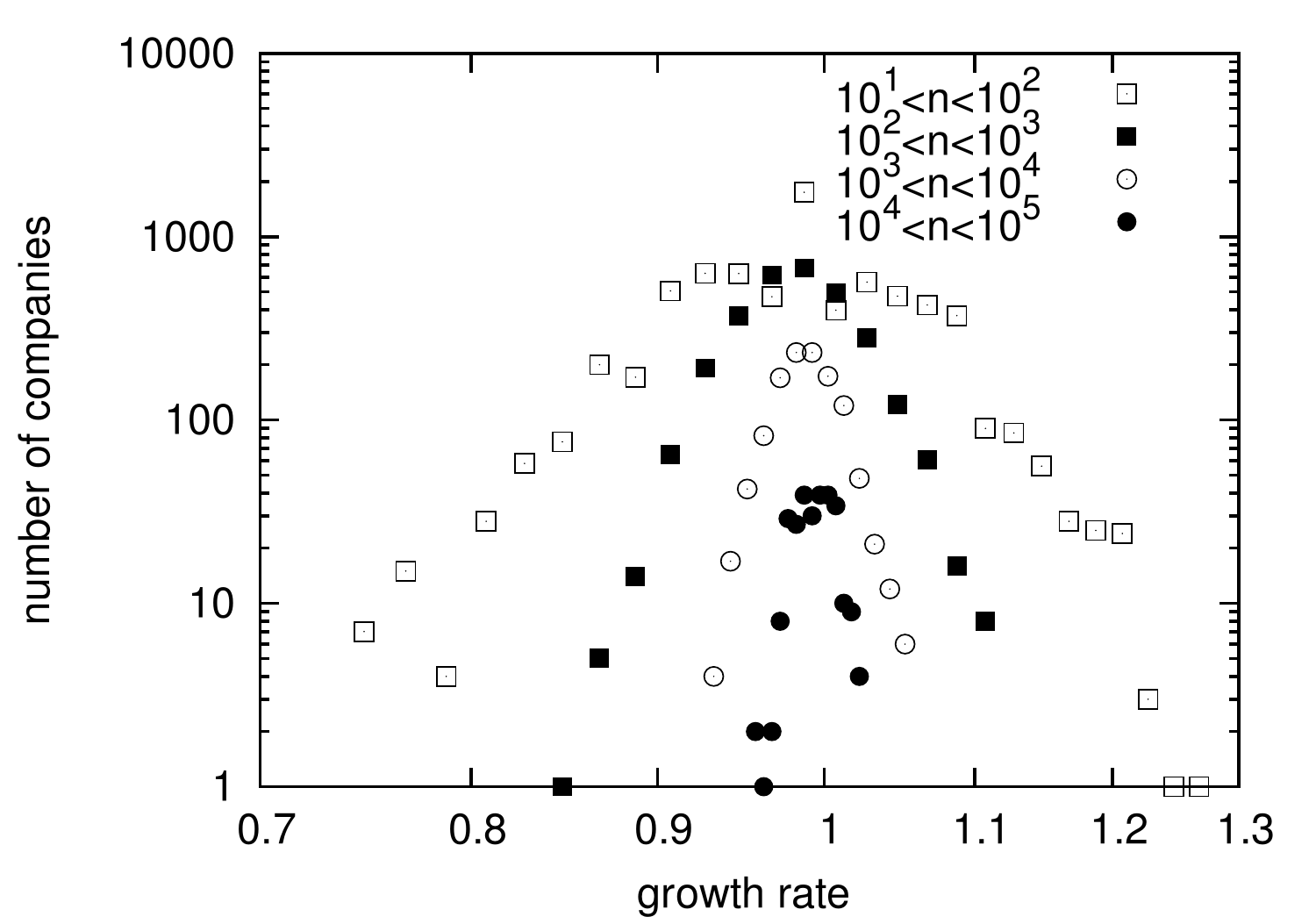}\caption{}\end{subfigure}
\begin{subfigure} [b]{0.45\textwidth}\includegraphics[angle=270, width =\textwidth]{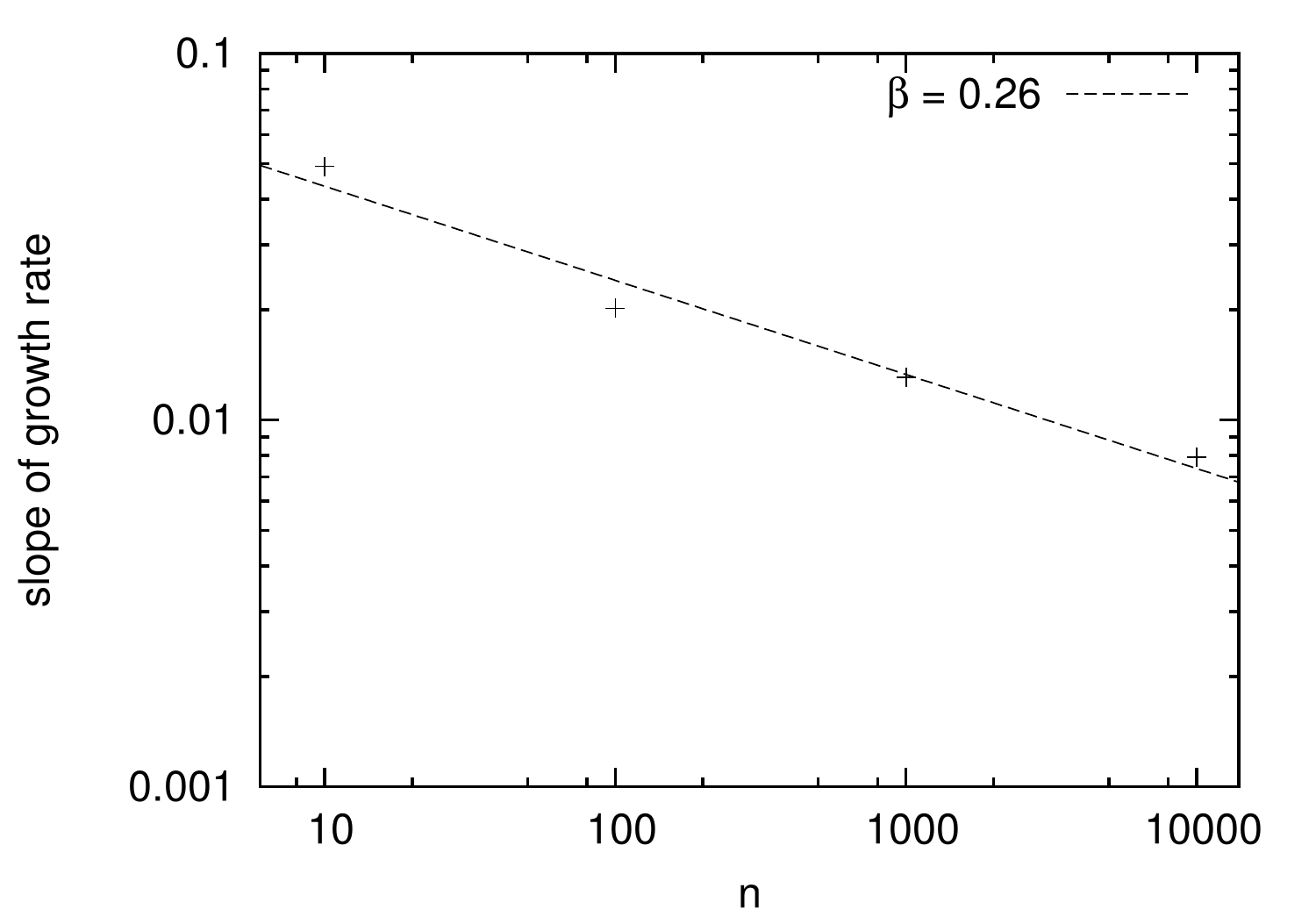}\caption{}\end{subfigure}
 	\caption{ (a) Growth rate distribution where companies are clustered into size bins (system with $10^5$ companies and $10^7$ workers), (b) scaling exponent determined from the slopes in (a). It yields $\beta$ = $0.26$, in agreement with the scaling exponent $\sigma\propto n^{-0.25}$ of the Gaussian growth rate probability densities with which the system has been simulated.}
	\label{fig:85cBIGscaling}
\end{figure}

The results of simulations shown in figures \ref{fig:85cBIGsim} and \ref{fig:85cBIGscaling} are, as expected, in between pure multiplicative noise $\beta$ = $0$ (figure \ref{fig:pure_multiplicative_noise}) and the presented model with scaling exponent $\beta $ = $0.5$ (figure \ref{fig:78aBIGgrowth}). 
The counter-cumulative size distribution on log-log-scale is closer to a Zipf law (i.e. $\alpha=1$) than the one from the presented model. In contrast, the slopes of the tent-shaped growth rate distribution appear less linear than for $\beta $ = $0.5$.

The strength of the hypothesis of Gaussian $\mathcal{G}(g|n)$ is that it would allow the explanation of heavy-tailed growth fluctuations as a collective phenomenon on aggregate level, without having to assume them on firm level, as e.g. \cite{schwarzkopf2010explanation}. Even if a Laplacian $\mathcal{G}(g|n)$ is assumed, as many authors do, the presence of a scaling exponent $\beta $ $\neq$ $0$ does not guarantee a power law for the size distribution. As the simulation in  figures \ref{fig:85cBIGsim} and \ref{fig:85cBIGscaling} show (which were not simulated with the presented model), this conclusion is independent of the rationale of the model as presented in section \ref{sec:model}.
\section{Comparison to other models}\label{sec:comparison_marsili}
The model can be compared to two existing models which also exhibit $\beta$ = $0.5$. The first is the city formation model of Marsili and Zhang \cite{marsili1998interacting}. They study two scenarios, of which one corresponds to pure multiplicative noise ($\beta$ = $0$) and yields a power law for the city size distribution, and one (which they term linear case) in which the growth rate standard deviation has scaling exponent $\beta =0.5$. For the latter scenario, Marsili and Zhang obtained an analytical expression for the size distribution as a function of the rank  $R$ of a city's size, $m(R)=m\cdot e^{-R+1}$, which is not a power law. No numerical result is shown. 

\begin{figure}[h!]
	\centering

\begin{subfigure}[b]{0.45\textwidth}
\includegraphics[angle=270, width =\textwidth]{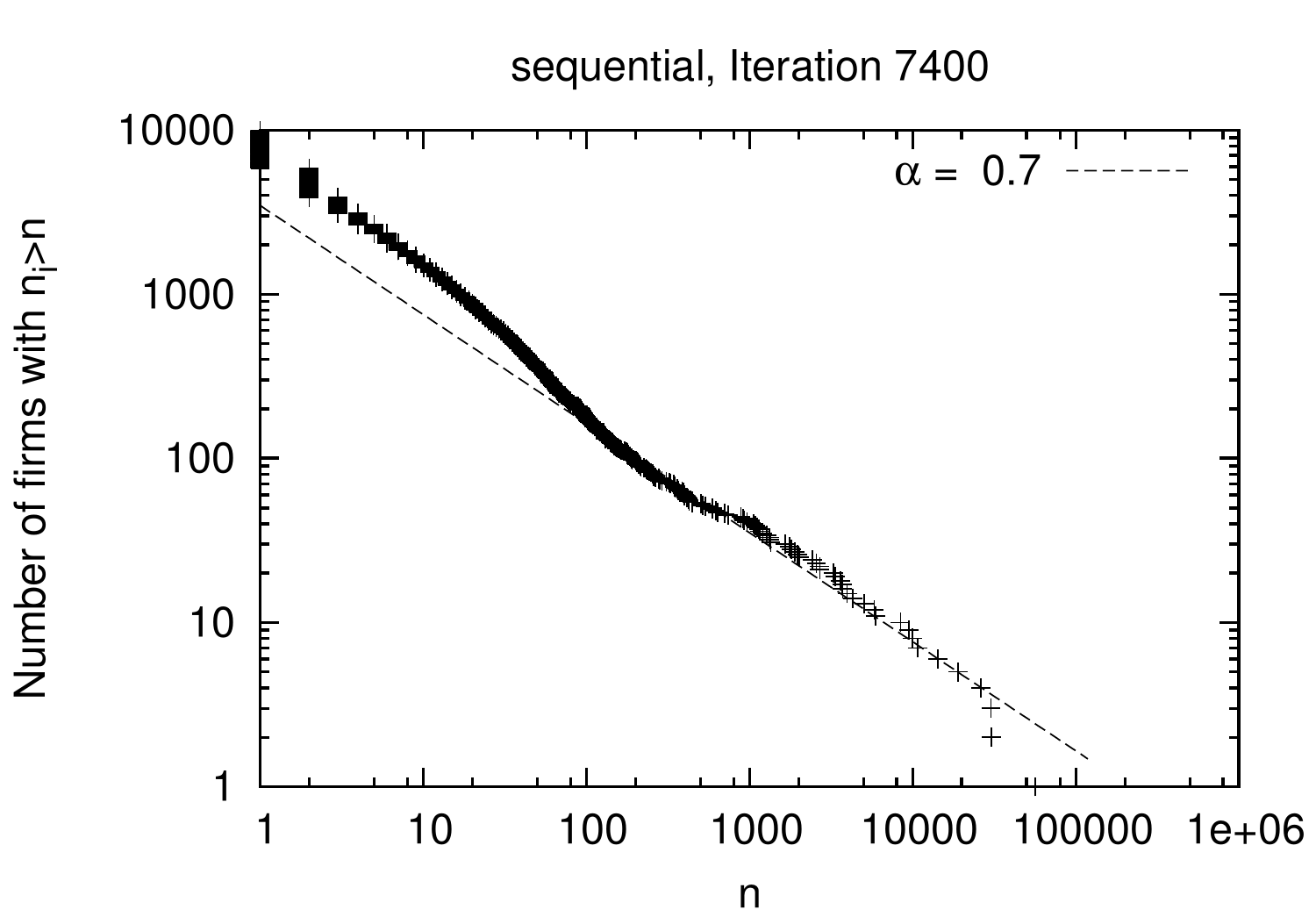}
\caption{}
\end{subfigure}
\begin{subfigure}[b]{0.45\textwidth}
\includegraphics[angle=270, width =\textwidth]{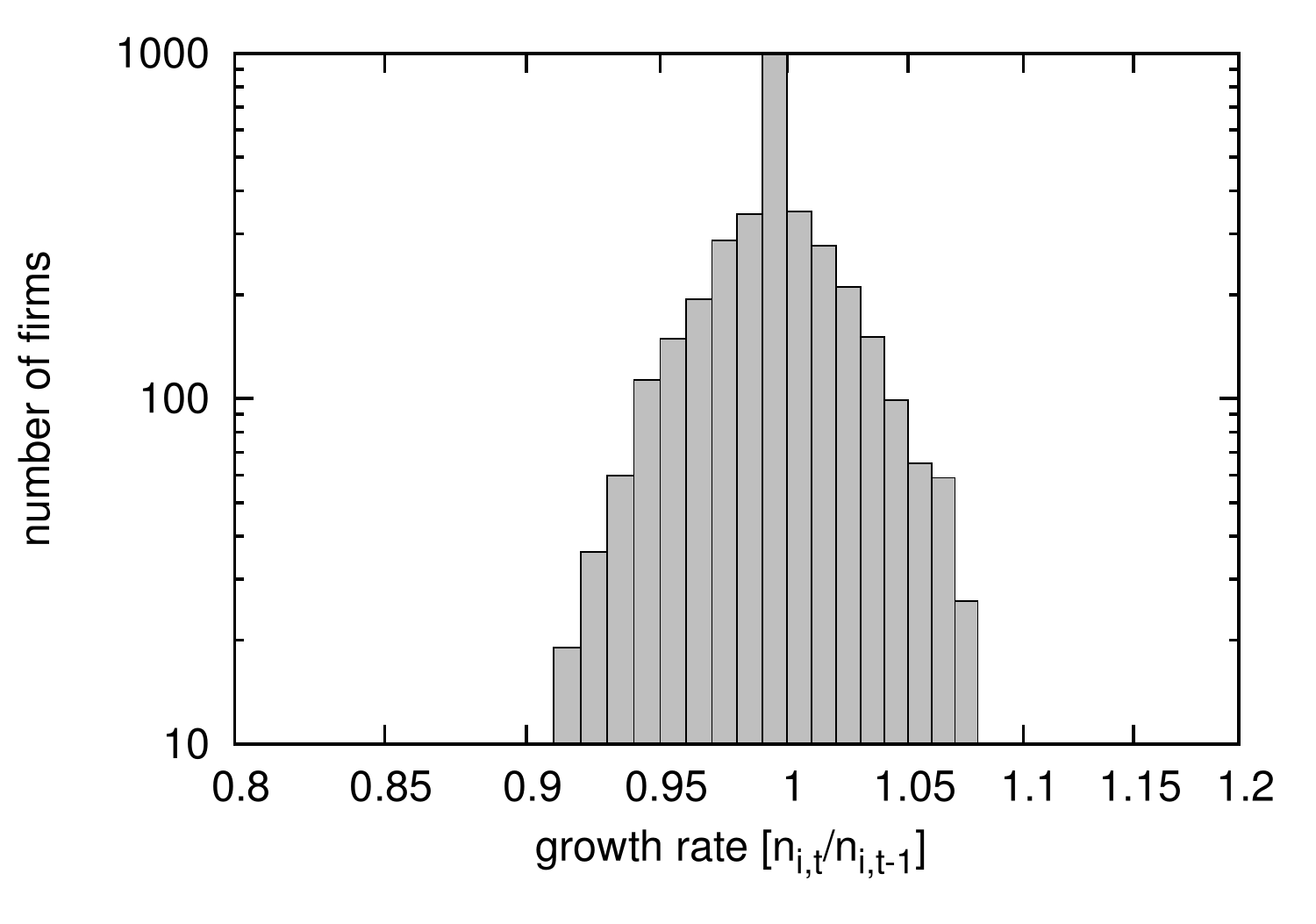}
\caption{}
\end{subfigure}
	\caption{(a) Counter-cumulative size distribution of city sizes from simulations of the linear model by Marsili and Zhang \cite{marsili1998interacting}, (b) the corresponding growth rate distribution  evaluated after 5\% of workers have changed city (company).}
	\label{fig:rank_size_marsili}
\end{figure}

The analogy to the model in this paper is easiest for case (i) where firms compete for available workforce, and all workers are employed. Then, workers can be considered to change company at each iteration, which is why firms grow and shrink. Marsili and Zhang's setting differs from our model in that city-dwellers do not move among cities all at the same time. 
It corresponds to a version of this model that is simulated in sequential update, a situation where workers drawn at random can change company, and the probability of joining a particular company is proportional to its size. 
We have simulated these sequential dynamics for comparison (see figure \ref{fig:rank_size_marsili}), since the authors do not show numerical results of their `linear case'. If the statistics of $\mathcal{G}(g|n)$ are calculated after a given number of changes at the level of workers, similar results to our tent-shaped $\mathcal{G}(g)$ are obtained, for the same reason as detailed in section \ref{sec:growth_rate}. A difference to the presented model is that a given city may change its size within the period over which the growth rate has been evaluated, so the probabilities of receiving or loosing a city-dweller may evolve during the movements of citizens that are all represented in \ref{fig:rank_size_marsili}. In contrast, in the model of this paper, these probabilities remain constant during one iteration. The results obtained for the size distribution are similar to the ones from the presented model, provided that $\mu$ was chosen sufficiently large. A conjecture is that the size distributions of the two models coincide, although it has been shown for other models that the choice of synchronous or asynchronous update does indeed influence the result \cite{huberman1993evolutionary}. Neither of the two models leads to a size distribution that can be fitted by exponential decay.

The second interesting model with $\beta$ = $0.5$ is the widely used model by Yule \cite{yule1925mathematical} and Simon \cite{simon1955class}, which also has binomial $\mathcal{G}(g|n)$ \cite{gabaix1999zipf}: if the constituent subunits of a firm were jobs, all existing jobs might double to two jobs with equal probability. In a particular time interval the variance of how many jobs doubled is narrower for large firms than for small firms, which may be described by a binomial distribution. (In addition, newly starting firms need to be taken into account). Yule's model leads to a Beta distribution, which exhibits a power law tail. This is due to the fact that the system is constantly growing, both in the number of employees and in the number of firms. However, it has been stated before that the power law is not found if these two assumptions are not satisfied \cite{krugman1996self}.
\section{Discussion}\label{sec:discussion}
Having presented the evolution of the size distribution and the growth rate distribution of this model (which has $\beta$ = $0.5$), as well as simulations of systems with additive noise ($\beta$ = $1$), multiplicative noise ($\beta =0$), and a system with $\beta =0.25$ for comparison, we return to some theoretical aspects of the model. Its dynamics can be described on three levels: The noise on the elementary (i.e. job/goods) level is the same for every element, which can double, vanish, or stay constant (see section \ref{sec:rounding_joboffer}). 

This elementary level allows for the calculation of size evolution of companies, which are the second level. On that level, the growth rate probability density is Gaussian, with a size-dependent variance. Because of this size dependency, integral (\ref{eq:growth_rate_integral}) becomes non trivial. The tent-shaped growth rate distribution of companies only holds at the aggregate system level, which is the third level. 

An analogy can be drawn to a physical system, where, due to long-range interactions, the statistics of an element and the statistics of the ensemble can differ. Beck and Cohen \cite{beck2003superstatistics} describe this by the term superstatistics, i.e. statistics of statistics, stating that in physical systems with fluctuations, the Boltzmann factor of the system is obtained by integrating the Boltzmann factors of every subsystem over their inverse temperatures. The analogy to the model presented here is the following: Instead of a Boltzmann factor, the quantity of interest is $\mathcal{G}(g|n)$, which depends on $n$. It is important to note that $\mathcal{G}(g|n)$ describes \textit{relative} fluctuations, which are normalized by $n$, whereas Boltzmann factors describe additive noise, i.e. 
 \textit{absolute} fluctuations. 
However, the concept of integrating over Boltzmann factors is the same as the integration over $n$-dependent growth rate variances in equation (\ref{eq:growth_rate_integral}). This $n$-dependence may also be seen as the result of long-range interactions: the hypothesis that every job is taken with the same probability implies that every open position interacts with every available employee. 

Physical systems with multiplicative noise, where the dynamics depend on the square of a Gaussian variable, exhibit the so-called q-exponential distribution \cite{beck2003superstatistics}\cite{biro2005power}, which can be derived from the extremality of the Rényi entropy. Multiplicative noise can be described as the interaction with a fluctuating external field. In contrast, in this model, the fluctuations come from competition for a limited resource, and exhibit different $n$-dependence, and different statistics as the distributions commonly found in physics. Further links of growth processes to the q-exponential distribution are presented in \cite{soares2005preferential,richmond2001power} and for the case of the tent-shaped growth rate distribution \cite{picoli2006scaling} and \cite{bettencourt2013hypothesis}.

\section{Conclusion}
\label{sec:conclusion_ch1}
In this paper a simple agent-based model has been introduced and analyzed, in which firm growth is the result of constraints in the markets, which can be the job market or the commodity goods market. Depending on whether or not firms spend their profits in the goods market, either workforce or aggregegate demand become scarce quantities in both markets respectively. These two scenarios have been simulated separately, but yield, as expected, very similar results. A matching algorithm, which is the same in the two markets, attributes this scarce quantity, and accounts for the growth dynamics of the system. Firm growth rates are size-dependent, where the standard deviations exhibit a scaling exponent $\sigma\propto n^{-0.5}$. 
In order to keep the size of the system constant, an additive term is needed, which here corresponds to the introduction of new firms whenever a firms has attained size $0$. 
The firm size distribution in the stationary state can be approximated by a power law of exponent $\alpha=0.7$, which is lower than that of a Zipf law with $\alpha=1$ found in data. This exponent is found in both scenarios and independently of profit margin $\mu$. This distribution has not yet been derived analytically, but is discussed in the context of existing results for Langevin systems with additive and multiplcative noise.

The second main result consists of the explanation of a tent-shaped growth rate probability density as a collective phenomenon. The presented model yields a growth rate probability density for firms that may be approximated by a Gaussian. Nevertheless, the aggregate growth rate pobability density of the system, for which there is empirical evidence, is tent-shaped.  This tent-shaped form is also found if firms are grouped into size bins, and a tent-shaped function is fitted to the growth rates of each bin, without the need to assume a Laplacian $\mathcal{G}(g|n)$. The central idea is to take firms' size distribution into account when calculating the growth rate distribution.  For comparison, simulations of a system  with size- dependent Gaussian multiplicative noise  ($\sigma\propto n^{-0.25}$) were carried out. Even for the latter case, which is in the range of empirical findings, the growth rate appears tent-shaped.

 \clearpage

\bibliography{thesis01}

\bibliographystyle{plain}
\begin{appendices}
 
\section{Alternative implementation: growth of independent subunits}\label{sec:rounding_joboffer}
The growth rate probability density of this model is binomial if there is a shortage of workforce (equation \ref{eq:binomial_dist}) or of purchasing power (equation \ref{eq:binomial_dist_goods}). For large $n$, this distribution can be approximated with a Gaussian distribution of the same variance. An alternative rounding method is now introduced which can be used for the determination of the job offer (case (i)) or the quantity to produce (case (ii)). It is detailed using the example of the job market.

The following setup yields a discrete Gaussian growth rate probability density even for small firms. Firms demand on average a quantity of workers $\hat n_i= n_i\,(1+\mu)$, which can be rounded towards integer values using the method introduced in equation \ref{eq:probabilistic_rounding}. Instead of rounding the quantity $\hat n_i$ to integers, the rounding can also be done at the level of individual positions: for every existing position $j$, the job offer $\hat j$ may be $1$ or $2$:
\begin{equation}
 \hat j=\begin{cases}
           2&\text{with probability } \mu\\ 
	   1&\text{with probability } (1-\mu)
          \end{cases}
\label{eq:indiv_rounding}
\end{equation}
Then, the job offer $\hat n_i$ is the sum of the job offers correponding to the positions of a firm.
\begin{equation}
 \hat n_i=\sum_{j=1}^{n_i}\hat j
\label{eq:sum_indiv_rounding}
\end{equation}
This $\hat n_i$ is the offer posted in the job market. Then, the aggregate job offer $\sum_i \hat n_i$ is collected. On average, it is $N(1+\mu)$, as with the standard rounding method. \paragraph{Combination with the allocation in the job market.} If available workforce $N_w$ is inferior to this offer (which is the case studied here), every open position has a probability 
\begin{equation}
 p=\frac{N_w}{\sum_i \hat n_i}\approx \frac{1}{1+\mu}
\end{equation}
of receiving a worker. This attribution on its own would yield a binomial constraint, depending on a firm's size, as stated in equation (\ref{eq:binomial_dist}). The growth rate has a cutoff at the upper value $(1+\mu)$, but firms can shrink to any size $\geq 0$. On the contrary, if firms determine their job offer via equations (\ref{eq:indiv_rounding}) and (\ref{eq:sum_indiv_rounding}), the number of received workers follows a symmetric distribution between $0$ and $2n$, if $n$ was the size of the firm in the previous timestep. Combining the probabilistic job offer (equation (\ref{eq:indiv_rounding})) with the binomial allocation of workers in the job market (equation \ref{eq:binomial_dist}), a single job has a certain probability to double, a certain probability to reproduce itself, and a certain probability to vanish:

 \begin{eqnarray}
p(j=2)=q=\frac{\mu}{(1+\mu)^2}\\
p(j=1)=p=\frac{1+\mu^2}{(1+\mu)^2}\\
p(j=0)=q=\frac{\mu}{(1+\mu)^2}\\
\label{eq:doubling_probability}
\end{eqnarray}

These probabilities are `reproduction probabilities' for single positions. For a firm of size $n$, the probabilities of receiving $k$ workers can be calculated out of these probabilities $p$ and $q$, in an analogous way as the coefficients of Pascal's triangle are found. It is indeed possible to establish a recursion relation for the coefficients $C$. The probability that a firm of size $n$ will have the size $k:= 2n-l$ in the following timestep is given by
\begin{equation}
 p(2n-l|n)=\sum_{j=0, n-l+2j>0, l-2j>0} \left(C(p^{l-2j-1}q^{n-l+2j})+C(p^{l-2j}q^{n-l-2j-1})\right)\,p^{l-2j}\,q^{n-l-2j}
\label{eq:recursion}
\end{equation}
In this derivation, the re-insertion of new firms has been neglected. Numerically, $G(g|n)$ is less noisy with this rounding method, compared to the case where firms offer precisely $(1+\mu)n_{i}$ jobs \footnote{Even in that case, $\lfloor(1+\mu)n_{i} \rfloor$ or $\lceil(1+\mu)n_{i} \rceil$ are offered, but no other values}. This rounding method is convenient because it yields a Gaussian $G(g|n)$ already for small firms.
\paragraph{}This method can also be applied to the production decision of firms. In combination with a shortage of purchasing power, a growth rate probability is derived in analogy to equation (\ref{eq:recursion}). An interpretation would be that for every sold good, a firm has a probability to sell 0, 1 or 2 in the following timestep, analogously to equations (\ref{eq:doubling_probability}). 

\begin{figure}[h!]
	\centering
\begin{subfigure}[b]{0.45\textwidth}

\includegraphics[angle=270, width =\textwidth]{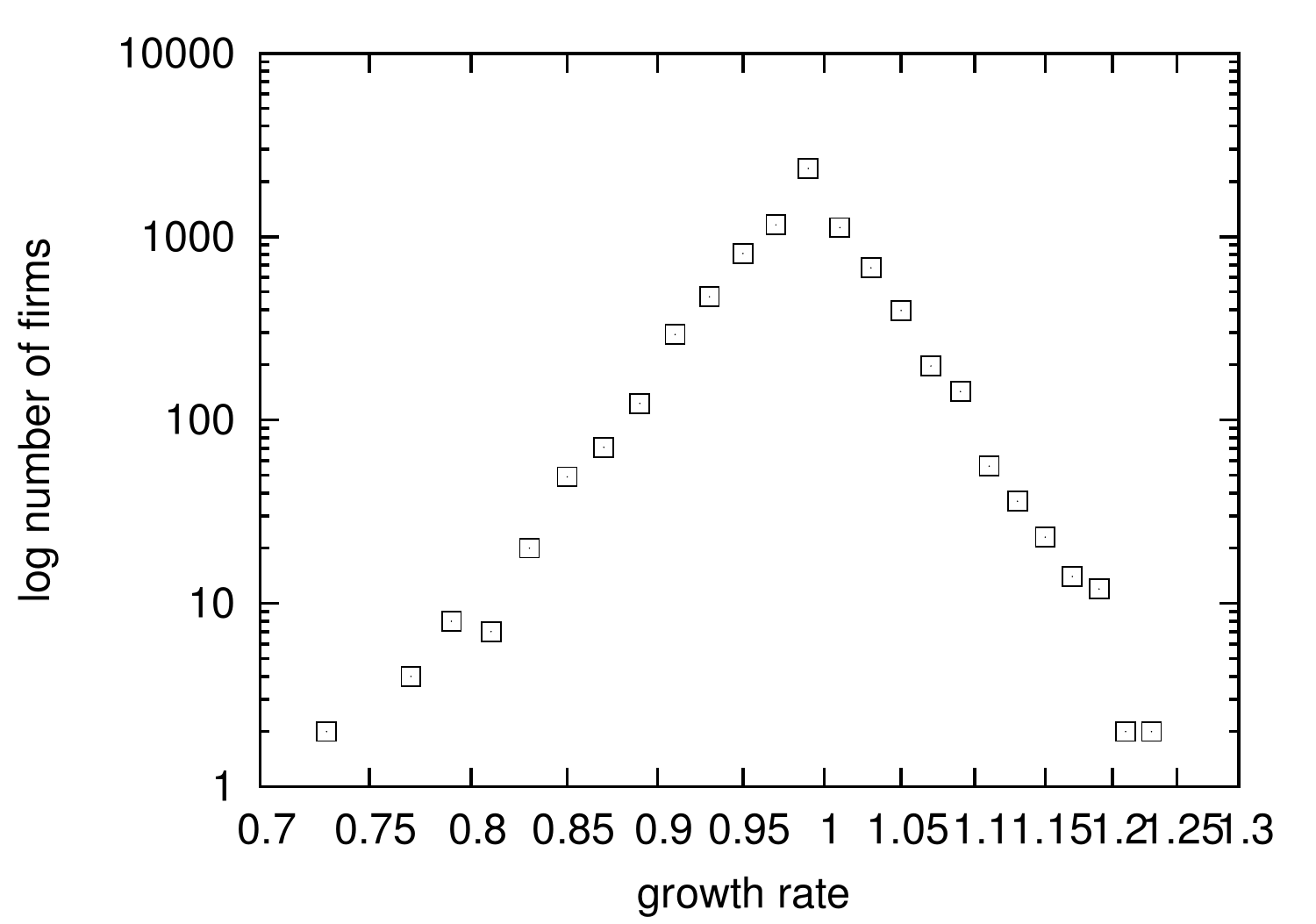}
\caption{}
\end{subfigure}
\begin{subfigure}[b]{0.45\textwidth}
\includegraphics[angle=270, width =\textwidth]{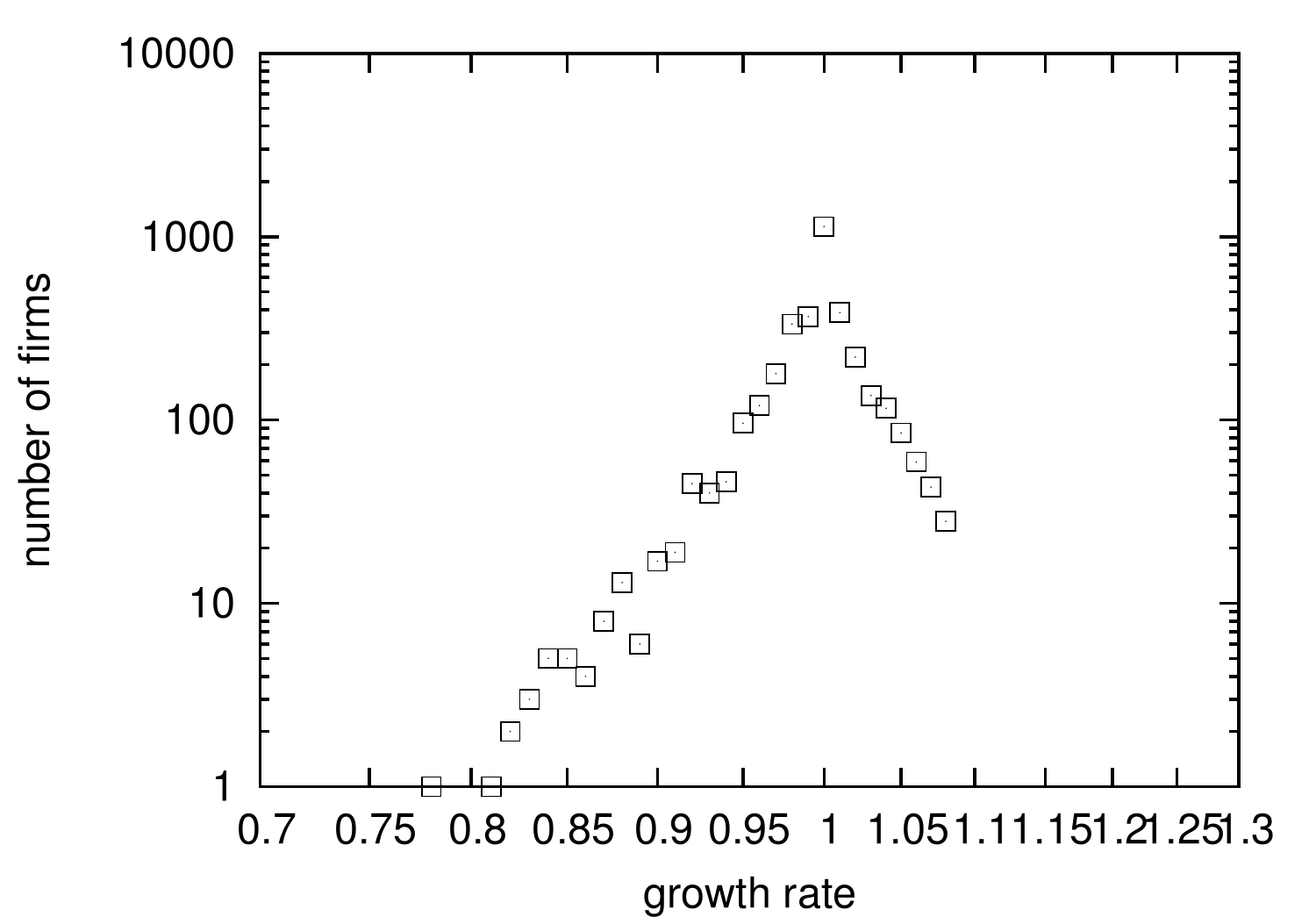}
\caption{}
\end{subfigure}
	\caption{(a) and (b): Two different rounding methods for $\mu=0.05$. Very small firms $q^s<10$ are removed from these statistics, since their growth rates cannot take continuous values, which will distort the statistics. For instance, a firm of size $2$ can only grow by $0,0.5, 1, 1.5$ and $2$, and since these small firms are numerous, peaks would be visible at these values.}

\end{figure}

\section{Numerical implementation}\label{sec:implementation}Some general technical details are given here. Parameters are found in the figure captions. 
\subsection{Rounding method}In equation (\ref{eq:q_hat}) $\hat q$ (and also $\hat n$) are not necessarily integer numbers and rounding is needed. In order to minimize rounding errors, the following method is introduced: 

Let $k_{\hat n} \in[0,1]$ be 
\begin{equation}
k_{\hat n}=[n(1+\mu)]- \lfloor n(1+\mu)\rfloor
\end{equation}
Rounding is then done using $k_{\hat n}$ as a probability:
\begin{equation}
\hat n=\begin{cases}
       \lceil n(1+\mu)\rceil&   \text{with probability }   k_{\hat n}\\
	  \lfloor n(1+\mu)\rfloor& \text{ with probability } 1-k_{\hat n}
          \end{cases}
\label{eq:probabilistic_rounding}
\end{equation}
This minimizes the rounding errors from discretization. This rounding method implies that it is possible for small firms to grow with a rate $g> (1+\mu)$: a firm of size $1$, which sold its entire production, demands $(1+\mu)$ workers in the next iteration, which may be rounded to $2$  with a probability of $k_{1+\mu}=\mu$.

\subsection{Additive term $\xi$ in scenarios (i) and (ii)}\label{sec:implementation_noise}
In scenario (i) where firms consume, and (ii) where firms do not consume, newly introduced firms affect the system in slightly different ways. In both scenarios, new firms contribute to the job offer $ \sum_j\hat n_j^{new}$
\begin{equation}
 \hat N= \sum_i \hat n_i + \sum_j\hat n_j^{new}
\end{equation}

\begin{itemize}
 \item In scenario (i), the stationary state is at full employment, and only a fraction $p=\frac{N_w}{\hat N}$ of positions will receive a worker. This $p$ will be slightly lower than $\frac{1}{1+\mu}$ whenever a new firm is started.
\item In scenario (ii), all positions are filled, so additional job offers cause an increase in workforce (i.e. decrease in unemployment) by $N^{new} \sum_j n_j^{new}$ workers. Therefore, the next production would be higher, and if iterated many times the system would tend towards full employment. In order to avoid this, the job offer of entrant firms $N^{new}$ is subtracted from existing firms. This is implemented such that every job offer of existing firms has the same chance of being eliminated. The result is that the job offer at iteration $t+1$ also equals the job offer at iteration $t$ when new firms are started. This means that the additive noise $\xi$, if considered for the entire system, has mean $0$, since it does not change the size of the system, but merely shifts some workers from existing firms with $n_i>0$ to firms with $n_i=0$. 
\end{itemize}

The latter method is  also used in the simulations of the similar model by Marsili and Zhang \cite{marsili1998interacting}, shown in figure \ref{fig:rank_size_marsili}. 
\section{Purely multiplicative and additive noise}
\begin{figure}[h!]
	\centering
\includegraphics[angle=270, width =0.7\textwidth]{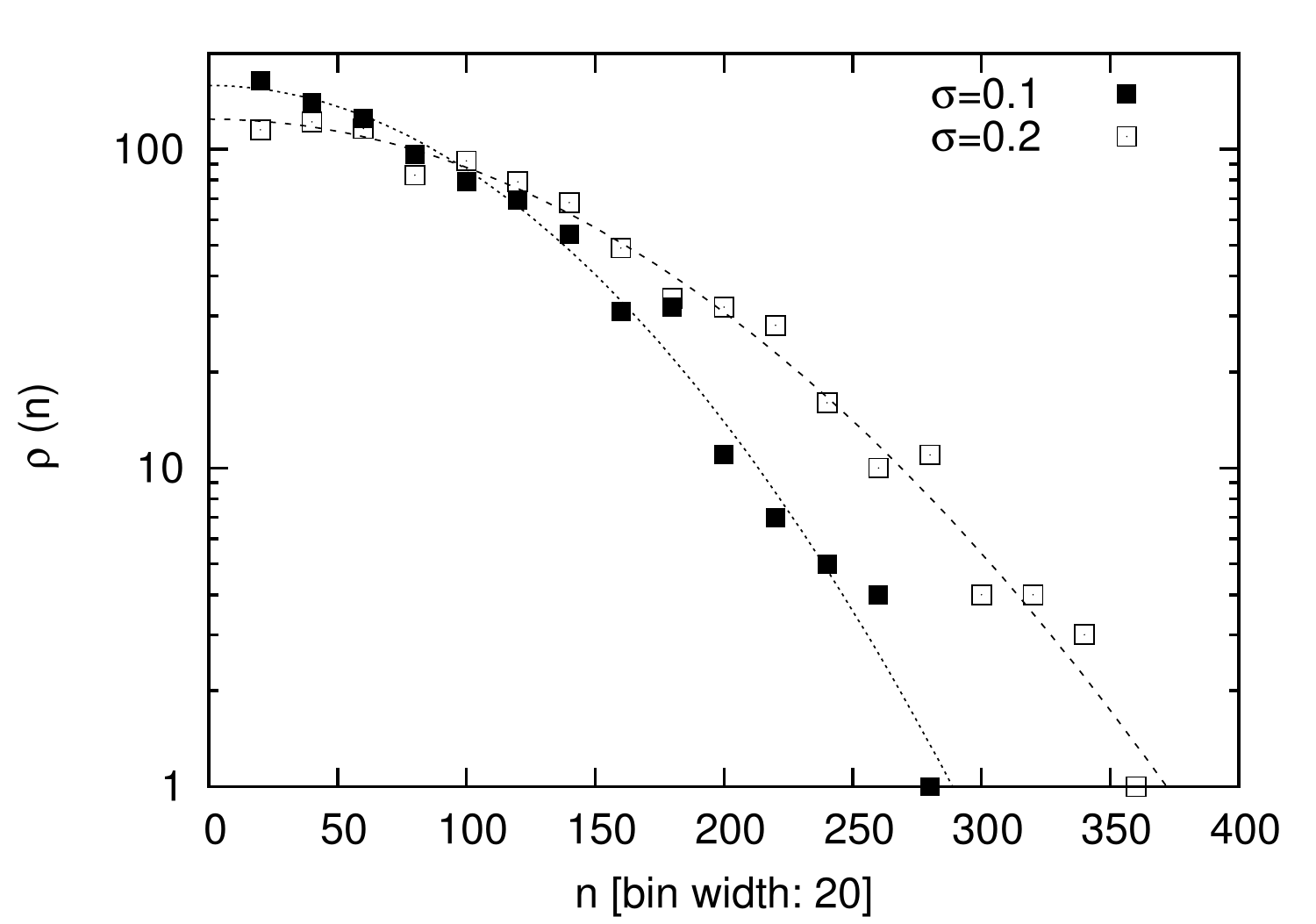}
 	\caption{Density of the size distribution of a system with purely additive uncorrelated Gaussian noise of mean zero, and absence of a multiplicative term. Simulations of two different standard deviations, on a log-linear scale. In this scale, a parabola corresponds to a Gaussian distribution. $N_w=10^5$, $N_f=10^3$. }
	\label{fig:pure_additive_noise}
\end{figure}

\begin{figure}[h!]
	\centering
\includegraphics[angle=270, width =0.7\textwidth]{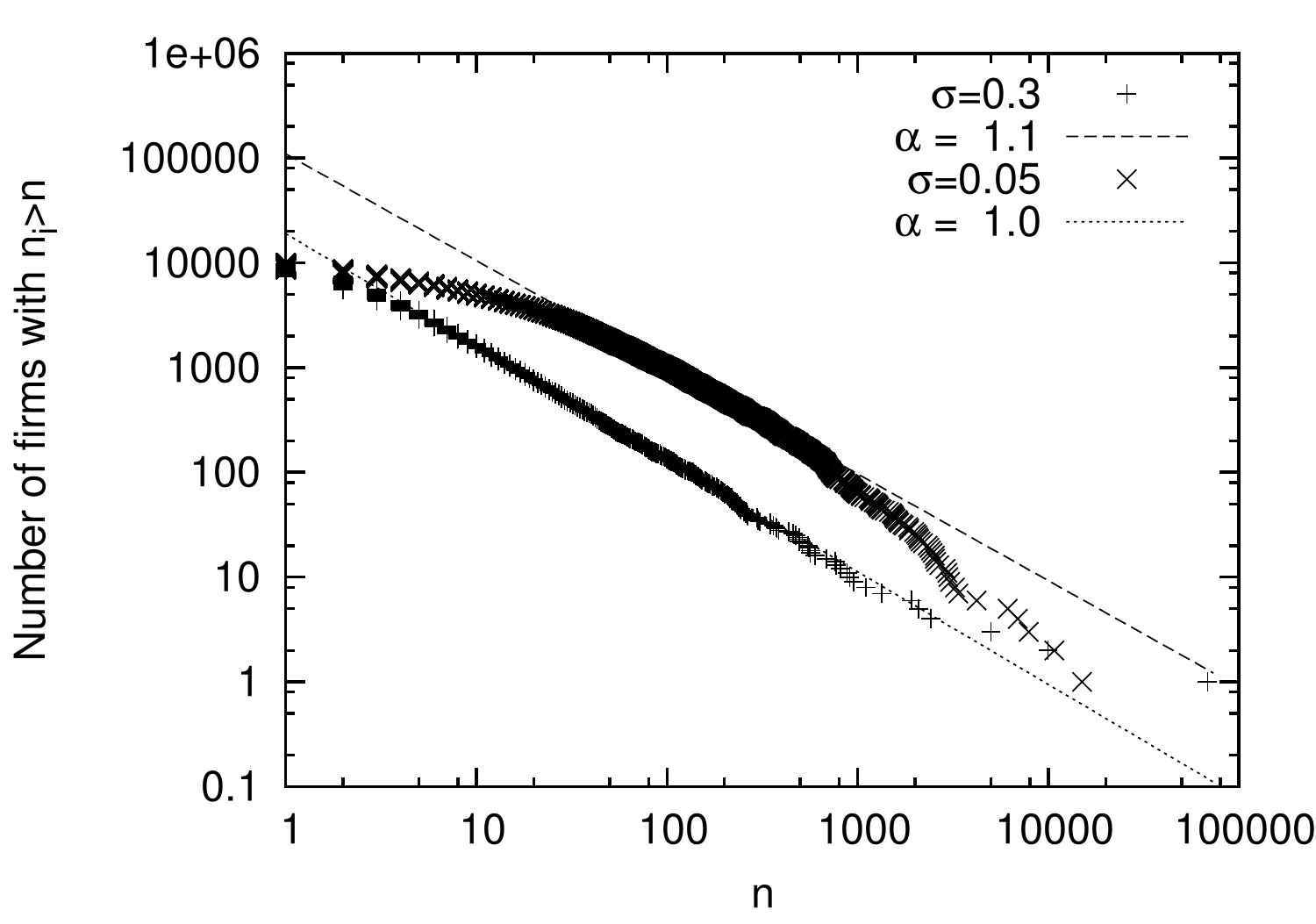}

	\caption{Numerical counter-cumulative size distributions of the Langevin equation (\ref{eq:langevin_discrete}), with multiplicative noise, which is uncorrelated and normally distributed. In this double logarithmic scale, a straight line corresponds to a power law. 
The smaller the variance of the noise distribution, the slower the convergence, and even after convergence the distribution remains concave. This is because rounding towards discrete values modifies the growth rate, such that it resembles additive noise. The smaller $\sigma$, the stronger this effect is. Results are from a system with $10^4$ firms and $5\cdot10^5$ workers, after $4000$ iterations. Whenever an element reached size $0$, it was replaced by one of average size $1.5$. The result is not sensititve to the size of re-initialisation.}
	\label{fig:pure_multiplicative_noise}
\end{figure}

Although the following two approximations do not correspond to the model introduced in section \ref{sec:model}, we simulated for comparison systems with purely additive noise (figure \ref{fig:pure_additive_noise}) and multiplicative noise (figure \ref{fig:pure_multiplicative_noise}). Only a small additional term $\xi$ has been present in order to keep the system at constant global size. Since the elements (firms) are discrete, rounding towards discrete values has the role of additive noise.

\end{appendices}

\end{document}